%% file: DISORDER.tex
\documentclass[11pt]{article} 

\usepackage{times}
\usepackage{amsmath}
\usepackage{amsthm}
\usepackage{amsfonts}
\usepackage{color}
\usepackage{algorithm}
\usepackage[noend]{algorithmic}
\usepackage{graphicx}
\usepackage{xcolor}
\usepackage{psfrag}
\usepackage{natbib}
\usepackage{epstopdf}
\usepackage{subfigure}
\usepackage{fullpage}
\usepackage{hyperref}
\usepackage{authblk}
\usepackage{multirow}
\usepackage{relsize}
\usepackage{overpic}
\usepackage{mdframed}
\usepackage{physics}
\usepackage{soul}
\usepackage{cancel}
\usepackage[labelfont=bf]{caption}
\usepackage[usestackEOL]{stackengine}
\usepackage[customcolors]{hf-tikz}
\usepackage{transparent}
\usepackage[many]{tcolorbox}
\newtcolorbox{crossfig}{blank,breakable,parbox=false,
  overlay={\draw[red,line width=5pt] (interior.south west)--(interior.north east);
    \draw[red,line width=5pt] (interior.north west)--(interior.south east);}}

\tcbsubskin{mycross}{empty,overlay=true}{frame code={\draw[red,line width=5pt] (frame.south west)--(frame.north east);\draw[red,line width=5pt] (frame.north west)--(frame.south east);},skin first=mycross,skin middle=mycross,skin last=mycross }

\tcbsubskin{mycrosstext}{empty,breakable,overlay=true}{frame code={\draw[red,line width=5pt] (frame.south west)--(frame.north east);\draw[red,line width=5pt] (frame.north west)--(frame.south east);},skin first=mycross,skin middle=mycross,skin last=mycross }

\graphicspath{{Figs/}}

\newcommand\argmin{\ensuremath{\operatornamewithlimits{argmin}}}

\newcommand\erfc{\ensuremath{\operatornamewithlimits{erfc}}}
\newcommand\ber{\fontfamily{cmr}\selectfont}

\title{Motion corrected MRI with DISORDER: Distributed and Incoherent Sample Orders for Reconstruction Deblurring using Encoding Redundancy}

\date{}
\author[1,2]{L Cordero-Grande}
\author[2]{G Ferrazzi}
\author[1,2]{RPAG Teixeira}
\author[1,3]{\\J O'Muircheartaigh}
\author[1,2]{AN Price}
\author[1,2]{JV Hajnal}
\affil[1]{Centre for the Developing Brain, School of Biomedical Engineering and Imaging Sciences, King's College London, King's Health Partners, St Thomas' Hospital, London, SE1 7EH, UK}
\affil[2]{Biomedical Engineering Department, School of Biomedical Engineering and Imaging Sciences, King's College London, King's Health Partners, St Thomas' Hospital, London, SE1 7EH, UK}
\affil[3]{Department of Forensic and Neurodevelopmental Sciences, Institute of Psychiatry, Psychology and Neuroscience, King's College London, King's Health Partners, 16 De Crespigny Park, London, SE5 8AF, UK}
\affil[ ]{\newline\small{lucilio.cordero\_grande@kcl.ac.uk,ferrazzi.giulio@gmail.com\\\{rui.teixeira,jonathanom,anthony.price,jo.hajnal\}@kcl.ac.uk}}

\definecolor{naranja}{rgb}{1.0,0.5,0}
\definecolor{violeta}{rgb}{0.5,0,1}
\definecolor{verde}{rgb}{0.75,1,0.75}
\definecolor{rojo}{rgb}{1,0.75,0.75}
\definecolor{azul}{rgb}{0.75,0.75,1}
\definecolor{dverde}{rgb}{0,0.5,0}
\definecolor{dlverde}{rgb}{0.25,0.75,0.25}
\definecolor{drojo}{rgb}{0.5,0,0}
\definecolor{dazul}{rgb}{0,0,0.5}
\definecolor{naranjal}{rgb}{1.0,0.85,0.7}

\tikzstyle{nosep}=[inner sep=0pt, outer sep=0pt]

\hfsetfillcolor{naranjal}
\hfsetbordercolor{white}

\def\figures{}

\begin{document}

\maketitle

\input{0abstract}
\input{1introduction}
\input{2theory}
\input{3methods}
\input{4results}
\input{5discussion}
\input{6conclusion}

\input{acknowledgments}

\clearpage

\bibliographystyle{apalike}
\bibliography{DISORDER}

\clearpage

\appendix
\input{7support}

\end{document}

%% file: 0abstract.tex
\begin{abstract}

{\bf \noindent Purpose:} To enable rigid-body motion tolerant parallel volumetric magnetic resonance imaging by retrospective head motion correction on a variety of spatio-temporal scales and imaging sequences.

{\bf \noindent Theory and methods:} Tolerance against rigid-body motion is based on distributed and incoherent sampling orders for boosting a joint retrospective motion estimation and reconstruction framework. Motion resilience stems from the encoding redundancy in the data, as generally provided by the coil array. Hence, it does not require external sensors, navigators or training data, so the methodology is readily applicable to sequences using 3D encodings.

{\bf \noindent Results:} Simulations are performed showing full inter-shot corrections for usual levels of in-vivo motion, large number of shots, standard levels of noise and moderate acceleration factors. Feasibility of inter- and intra-shot corrections is shown under controlled motion in-vivo. Practical efficacy is illustrated by high quality results in most corrupted of $208$ volumes from a series of $26$ clinical pediatric examinations collected using standard protocols.

{\bf \noindent Conclusion:} The proposed framework addresses the rigid motion problem in volumetric anatomical brain scans with sufficient encoding redundancy which has enabled reliable pediatric examinations without sedation.

\end{abstract}


%% file: 1introduction.tex
\ifdefined\figures
\section{Introduction}
\else
\section*{Introduction}
\fi

\label{sec:INTR}

Tolerance against motion is desirable in magnetic resonance imaging (MRI). This includes brain MRI, where significant motion-induced image degradation prevalence has been documented~\citep{Andre15} and high resolution imaging quality may be compromised by head motion~\citep{Budde14}. Rigid-body MRI motion correction~\citep{Zaitsev15,Godenschweger16} can be tackled via prospective or retrospective techniques. Prospective techniques~\citep{Maclaren13} compare advantageously in terms of spin-history, dephasing confounders or k-space density guarantees. Particularly, optical tracking systems have been proposed for head motion estimation, with corrections showing impressive accuracy and latency~\citep{Schulz12}. However, prospective methods require additional hardware and/or scanner modifications, often involving intrusive markers attached to the subject. In addition, satisfactory corrections may not always be possible due to unpredictability or complexity of motion, or maximized sampling efficiency requirements. Retrospective techniques may facilitate scanning or improve prospective results~\citep{Aksoy12}, particularly for 3D encodings, where spin-history is less of a problem, and when using non-linear reconstruction paradigms~\citep{Adcock14} to deal with non-homogeneous sampling density after motion.

Motion compensation is strongly dependent on motion estimation from the measured information. Some methods have proposed the use of navigators, where surrogate motion-sensitive information is interleaved with the main acquisition and correction is applied either prospective or retrospectively~\citep{Tisdall16,Johnson16,Gallichan17}. Due to variability in time requirements for different MRI sequences, application of a given navigator is usually limited to a specific sequence type. Furthermore, particular care has to be taken to prevent spin-history or saturation and, sometimes, scanning efficiency may be compromised. Alternatively, sequences can be constructed with relative resilience to motion or, similarly, sampling schemes can be designed to function as implicit navigators. This is the case for spiral and radial trajectories~\citep{Bammer07,Anderson13,Pipe14}, where temporally distributed low resolution information is used for motion estimation, with retrospective corrections usually grounded on an intermediate reconstruction of fully formed images for each motion state, often involving non-linear methods. Finally, other approaches have explored the redundancy of the information sensed by parallel MRI to detect and discard localized inconsistencies in k-space measurements~\citep{Bydder02}, usually requiring prior image models to limit noise amplification and improve inconsistency detection~\citep{Samsonov10}.

Building on models of MRI acquisition in the presence of motion~\citep{Batchelor05,Bammer07}, some methods have proposed formulations for motion estimation from the k-space that do not require navigators~\citep{Odille08,Loktyushin13,Cordero-Grande16}. Our previous work~\citep{Cordero-Grande16} introduced a data-driven reconstruction method for retrospective multi-shot rigid-body motion correction or \emph{aligned reconstruction} taking advantage of the encoding redundancy in the measured data. Performed simulations showed that the ability to solve the aligned reconstruction problem is strongly sensitive to the k-space encoding order, which suggested that opportunities exist to maximize the sensitivity to motion by appropriate sampling order designs. Consequently, in this paper we introduce the Distributed and Incoherent Sample Orders for Reconstruction Deblurring using Encoding Redundancy (DISORDER) framework as a flexible way to correct for head motion on a variety of spatio-temporal scales and imaging contrasts by optimizing the \emph{sample orders} for k-space coverage. In addition, we propose some technical refinements to the aligned reconstruction formulation and extend the simulation domain. The technique is implemented on a $3\,\mbox{T}$ scanner and tested on controlled motion scans and pediatric examinations including magnetization-prepared rapid acquisition gradient echo (MP-RAGE), fast spin echo (FSE), fluid attenuated inversion recovery (FLAIR), spoiled gradient echo (SPGR), and balanced steady-state free precession (bSSFP) sequences. A \textsmaller{\textsc{MATLAB}} implementation to reproduce the experiments is made available at \url{https://github.com/mriphysics/DISORDER/releases/tag/1.1.0}.

%% file: 2theory.tex
\ifdefined\figures
\section{Theory}
\else
\section*{Theory}
\fi

\label{sec:THEO}

\ifdefined\figures
\subsection{Aligned reconstruction}
\else
\subsection*{Aligned reconstruction}
\fi

\label{sec:ALSE}

Assuming whitened measurement noise~\citep{Pruessmann01}, the aligned reconstruction for parallel volumetric imaging can be formulated as:
\begin{equation}
\label{ec:GEFO}
(\hat{\mathbf{x}},\hat{\boldsymbol{\theta}})=\displaystyle\argmin_{\mathbf{x},\boldsymbol{\theta}}r_{\mathbf{x},{\boldsymbol{\theta}}}=\displaystyle\argmin_{\mathbf{x},\boldsymbol{\theta}}\|\mathbf{A}\boldsymbol{\mathcal{F}}\mathbf{S}\mathbf{T}_{\boldsymbol{\theta}}\mathbf{x}-\mathbf{y}\|_2^2,
\end{equation}
where $\mathbf{x}$ is the image to be reconstructed, $\boldsymbol{\theta}$ are the motion parameters, $r$ is the loss function, $\mathbf{y}$ is the measured k-space data, $\mathbf{T}$ is a set of rigid motion transformations, $\mathbf{S}$ are the coil sensitivities, $\boldsymbol{\mathcal{F}}$ is the discrete Fourier transform (DFT), and $\mathbf{A}$ is a sampling mask. We are interested in reconstructing a 3D image of size $V=V_1V_2V_3$ with $V_d$ the number of voxels along dimension $d$ from $\displaystyle N=C\sum_{m=1}^{M}E_m$ $C$-element coil array samples of a discretized k-space grid of size $K$. $E_m$ denotes the number of samples within \emph{segment} $m$ and $M$ is the number of segments in the sequence, with each segment associated to a specific motion state. Detailed information about the terms in Eq.~\eqref{ec:GEFO} can be found in~\cite{Cordero-Grande16}. Here we provide a brief description of their structure:
\begin{itemize}
	\item $\mathbf{y}$ is a $N\times 1$ vector.
	\item $\mathbf{A}$ is a $N\times KMC$ block matrix comprising submatrices of size $E_m\times K$ whose entries take the value $1$ if the sample $e$ of the segment $m$ corresponds to the k-space location indexed by $k$ and $0$ otherwise.
	\item $\boldsymbol{\mathcal{F}}$ is a $KMC\times VMC$ block diagonal matrix comprising submatrices of size $K\times V$ representing 3D DFT's with applied k-space sampling.
	\item $\mathbf{S}$ is a $VMC\times VM$ block matrix comprising diagonal submatrices of size $V\times V$ whose diagonal elements correspond to the spatial sensitivity of the coil $c$.
	\item $\mathbf{T}$ is a $VM\times V$ block matrix comprising unitary~\citep{Unser95} submatrices of size $V\times V$ corresponding to the 3D rigid transformation modeling the motion state $m$ by three translations and three Euler rotation angles codified in the parameter vector $\boldsymbol{\theta}_m$.
	\item $\mathbf{x}$ is a $V\times 1$ vector.
\end{itemize}

Eq.~\eqref{ec:GEFO} is a separable nonlinear least squares problem~\citep{Gan18,Herring18}. We confront it by iteratively addressing the subproblems:
\begin{equation}
\label{ec:GEFD}
\begin{split}
\hat{\mathbf{x}}^{(i+1)}=&\displaystyle\argmin_{\mathbf{x}}\|\mathbf{A}\boldsymbol{\mathcal{F}}\mathbf{S}\mathbf{T}_{\hat{\boldsymbol{\theta}}^{(i)}}\mathbf{x}-\mathbf{y}\|_2^2\\
\hat{\boldsymbol{\theta}}^{(i+1)}=&\displaystyle\argmin_{\boldsymbol{\theta}}\|\mathbf{A}\boldsymbol{\mathcal{F}}\mathbf{S}\mathbf{T}_{\boldsymbol{\theta}}\hat{\mathbf{x}}^{(i+1)}-\mathbf{y}\|_2^2.
\end{split}
\end{equation}
The first subproblem, reconstructing the image $\mathbf{x}$ in the presence of rigid motion~\citep{Batchelor05}, can be solved by conjugate gradient (CG)~\citep{Pruessmann01}. As for the second, the solution must null the gradient of the objective function against the motion parameters~\citep{Cordero-Grande16}, which is tackled by a Levenberg-Marquardt (LM) algorithm using a simplified Jacobian~\citep{Ruano91}. A natural initialization is a zero-motion condition $\hat{\boldsymbol{\theta}}^{(0)}=\mathbf{0}$, so the first step corresponds to a standard sensitivity encoding (SENSE) reconstruction. Further in this paper, we describe how to temporally arrange the k-space samples into segments to improve the aligned reconstruction convergence.

\ifdefined\figures
\subsection{DISORDER sampling}
\else
\subsection*{DISORDER sampling}
\fi

\label{sec:SATE}

We focus on Cartesian 3D k-space grids with uniform sampling as sketched in Fig.~\ref{fig:VOEN}. Fig.~\ref{fig:VOEN}a shows $K_1=4$ collected samples after the first readout or profile in the $k_1$ direction. Fig.~\ref{fig:VOEN}b shows the first segment, in this example corresponding to the full acquisition of the $k_1$-$k_2$ plane. Fig.~\ref{fig:VOEN}c shows that segments can be used to define an ordered partition of the $k_1$-$k_2$-$k_3$ grid. Due to short duration, we assume negligible motion during the readout and focus on the phase encode (PE) plane in Fig.~\ref{fig:VOEN}d. We define $E^{\text{PE}}_m$ as the number of profiles per segment, so $E^{\text{PE}}_m=E_m/K_1$, and hereinafter we adopt the replacement $E_m\leftarrow E^{\text{PE}}_m$.
\ifdefined\figures
\begin{figure}[!htb]
\begin{center}
\begin{overpic}[width=0.2495\textwidth,draft=false]{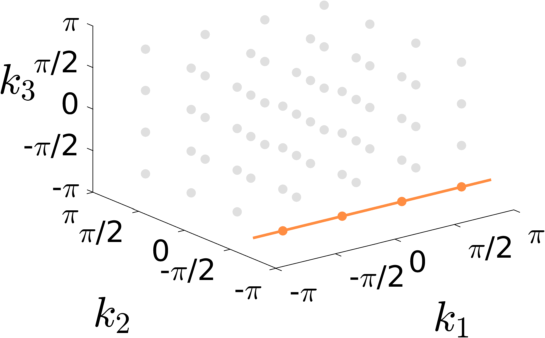}\put(0,0){\textbf{a)}}\end{overpic}
\begin{overpic}[width=0.2495\textwidth,draft=false]{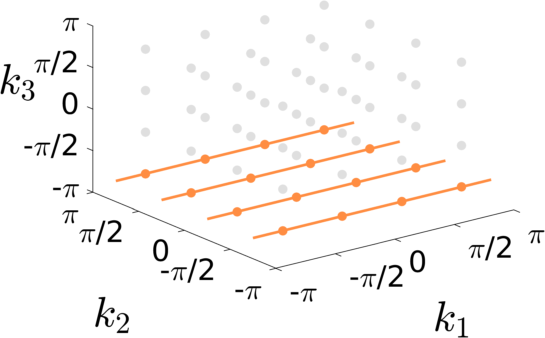}\put(0,0){\textbf{b)}}\end{overpic}
\begin{overpic}[width=0.2495\textwidth,draft=false]{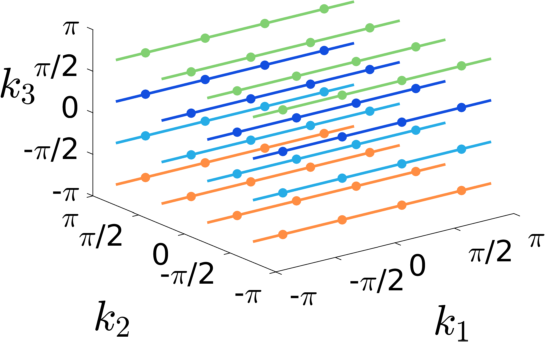}\put(0,0){\textbf{c)}}\end{overpic}
\begin{overpic}[width=0.2495\textwidth,draft=false]{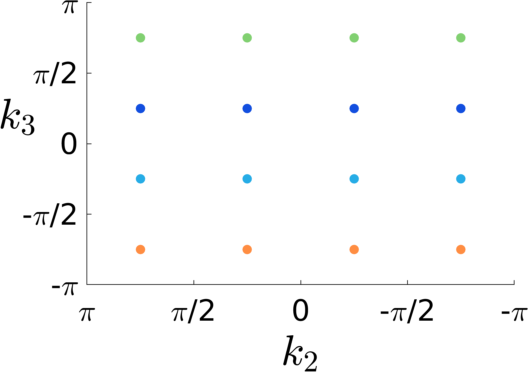}\put(0,0){\textbf{d)}}\end{overpic}
\end{center}\vspace{-5mm}
\input{caption1}
\end{figure}
\fi

By modifying the PE gradients before each readout it is possible to design different encoding or view orders. These can be defined as a temporally ordered set of profiles 
\ifdefined\figures
\newline
\fi
 $p\in\mathcal{P}=\{(k^{1,1}_2,k^{1,1}_3),\ldots,(k^{1,E_1}_2,k^{1,E_1}_3),\ldots,(k^{M,E_M}_2,k^{M,E_M}_3)\}$ with cardinality $\displaystyle P=|\mathcal{P}|=\sum_{m}E_m$. Fig.~\ref{fig:SAOR}a shows the first segment of a commonly used {\ber Sequential} ordering scheme. In this case, due to the partition definition, a segment includes two consecutive $k_3$ planes. Fig.~\ref{fig:SAOR}b, introduces the {\ber Checkered} traversal. First, a rectangular tiling of the PE plane is built using tiles of size $U_2\times U_3$ such that $U_2U_3=M$. Second, a spectral lexicographic order for the profiles within a tile $\mathcal{K}_U$ is defined by $\mathcal{K}_U\to\mathcal{M}_U=\{1,\ldots,M\}$, which can be extended to different tiles by translation. Third, interleaved segments are defined such that the same profile $m_u\in\mathcal{M}_U$ is used $\forall e\in\mathcal{E}=\{1,\ldots,E\}$, with $\mathcal{E}$ a temporally ordered set of tiles. Finally, the profile sequence to traverse each tile is defined by mapping from the set of profiles within a tile to a temporally ordered set of segments $\mathcal{M}_U\to\mathcal{M}_{T}=\{1,\ldots,M\}$ using an electrostatic repulsion criterion with periodic boundary conditions. Hence, a \emph{distributed} temporal coverage is guaranteed both for the whole spectrum and within each tile. This strategy aids the aligned reconstruction conditioning by reducing the chances for large uncovered spectral areas due to head rotation. Fig.~\ref{fig:SAOR}c presents the {\ber Random-checkered} modification. Tiles are built as in the {\ber Checkered} approach but segments are constructed by a random permutation of the tile elements drawn independently for each tile, so we have $m_{u_e}$. This guarantees a distributed coverage in probability and introduces some \emph{incoherence} among profiles within and across segments. Fig.~\ref{fig:SAOR}d shows a {\ber Random} view order where $\mathcal{P}$ is a random permutation of the profiles conforming the sequence. Note that, considering a free definition of segments, {\ber Sequential} and {\ber Random} schemes are particular cases of the {\ber Random-checkered} traversal respectively with tiling sizes of $1\times 1$ and $K_2\times K_3$. 

\ifdefined\figures
\begin{figure}[!htb]
\begin{center}
\begin{overpic}[width=0.2495\textwidth,draft=false]{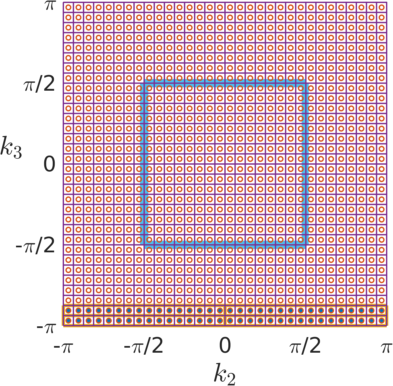}\put(8,0){\textbf{a)}}\end{overpic}
\begin{overpic}[width=0.2495\textwidth,draft=false]{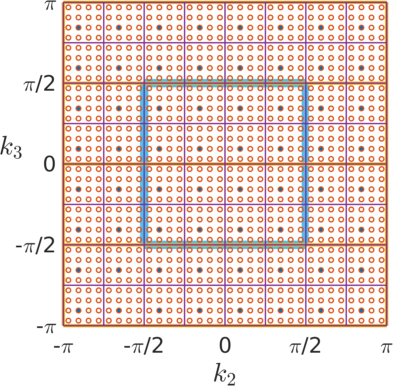}\put(8,0){\textbf{b)}}\end{overpic}
\begin{overpic}[width=0.2495\textwidth,draft=false]{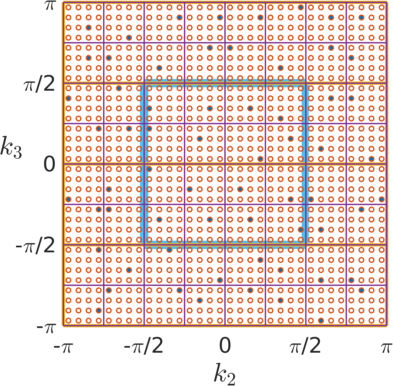}\put(8,0){\textbf{c)}}\end{overpic}
\begin{overpic}[width=0.2495\textwidth,draft=false]{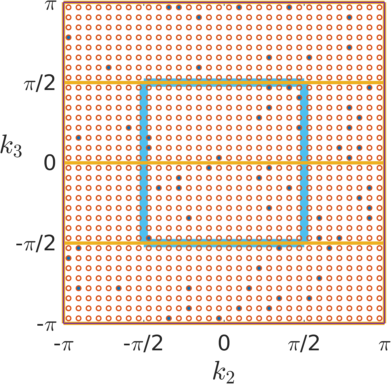}\put(8,0){\textbf{d)}}\end{overpic}
\end{center}\vspace{-5mm}
\input{caption2}
\end{figure}
\fi

View orders should preserve the contrast and the trajectory consistency. We establish the following differentiation:
\begin{itemize}
	\item \textbf{Non steady-state sequences} (MP-RAGE, FSE, FLAIR). They acquire a fraction of the k-space or \emph{shot} after each radiofrequency (RF) preparation. Thus, they induce a natural sampling partition where segments are in correspondence to shots. In addition, magnetic properties are not invariant for the different shot samples. Typically, middle samples within each shot cover the central area of the spectrum~\citep{Busse08}, which our orders can fulfill by jumping from tile to tile in a {\ber Zig-zag} manner. An example is shown in Fig.~\ref{fig:ECORA} using an elliptical sampling area. The {\ber Checkered} traversal produces regular segment patterns (first column), with neighboring colors maximally separated within the tile, whereas the {\ber Random-checkered} traversal produces non-regular patterns. Tiling orders generate smooth color transitions across the spectrum for all presented traversals (second column), which translates into smooth magnetic properties of the profiles.
	\item \textbf{Steady-state sequences} (SPGR, bSSFP). They produce a temporally stable magnetization after reaching the steady-state, usually facilitated by some preparatory dummy profiles, so the contrast becomes independent of the encoding order. For estimates attempted at the segment level, temporal resolvability of motion will increase with bigger tiling sizes $U_2U_3$. However, large jumps in the spectrum may induce inconsistencies due to eddy currents, especially for low repetition times, so a trade-off may be required. If the profiles are covered by the application of $M$ spectral \emph{sweeps}, analogously to~\cite{Tsao05}, eddy currents can be minimized by an {\ber Alternating zig-zag} tiling order where the traversal polarity is reversed for consecutive sweeps. This is illustrated in Fig.~\ref{fig:ECORB}. The segment structure (first column) matches that of Fig.~\ref{fig:ECORA} but for some minor differences in the {\ber Sequential} case due to smooth magnetization requirements in shot-based sequences~\citep{Busse08}. The {\ber Sequential} scheme guarantees a smooth passage through k-space (third and fourth columns). Although quicker k-space sweeps of our traversals imply larger $\dd k_2$ and $\dd k_3$ steps, these remain substantially lower than for the {\ber Random} order, which should limit the impact of eddy currents. Finally, the {\ber Alternating zig-zag} suppresses the undesirable spikes in the fourth column of Fig.~\ref{fig:ECORA}.
\end{itemize}
\ifdefined\figures
\begin{figure}[!htb]
\begin{center}
\begin{overpic}[width=0.246\textwidth,draft=false]{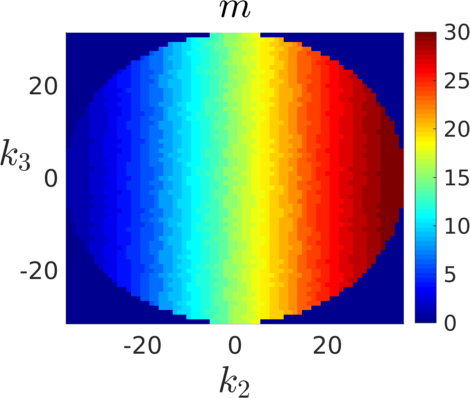}\put(0,0){\textbf{a)}}\end{overpic}
\begin{overpic}[width=0.246\textwidth,draft=false]{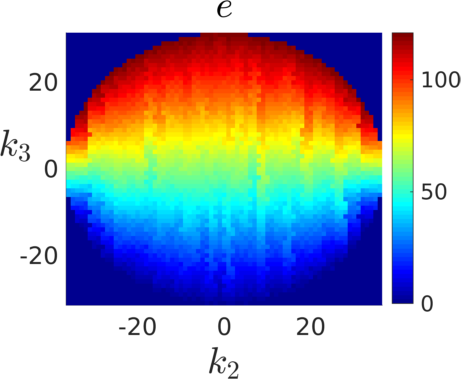}\end{overpic}
\begin{overpic}[width=0.246\textwidth,draft=false]{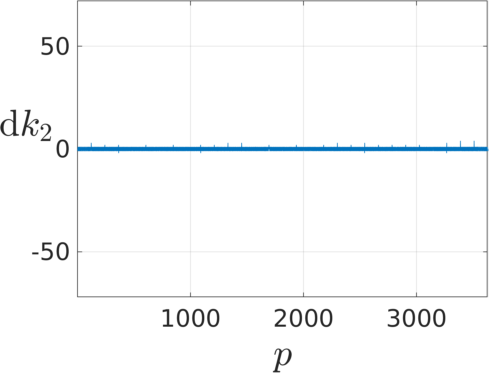}\end{overpic}
\begin{overpic}[width=0.246\textwidth,draft=false]{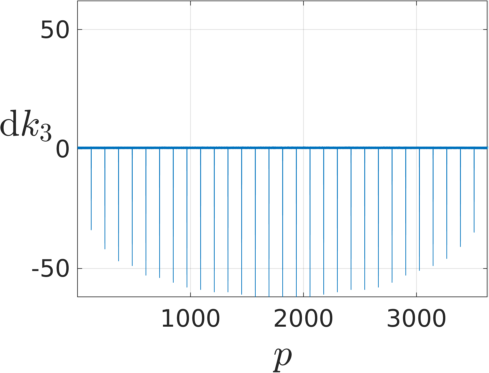}\end{overpic}\\\vspace{2mm}
\begin{overpic}[width=0.246\textwidth,draft=false]{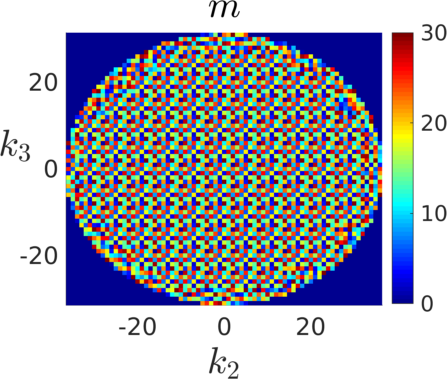}\put(0,0){\textbf{b)}}\end{overpic}
\begin{overpic}[width=0.246\textwidth,draft=false]{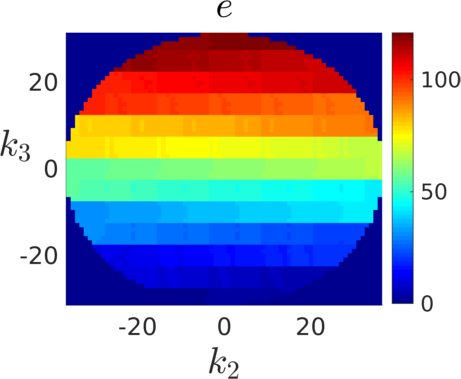}\end{overpic}
\begin{overpic}[width=0.246\textwidth,draft=false]{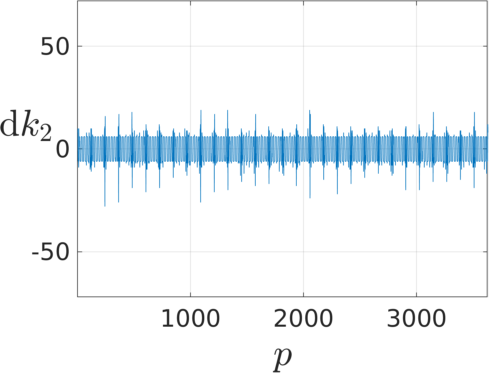}\end{overpic}
\begin{overpic}[width=0.246\textwidth,draft=false]{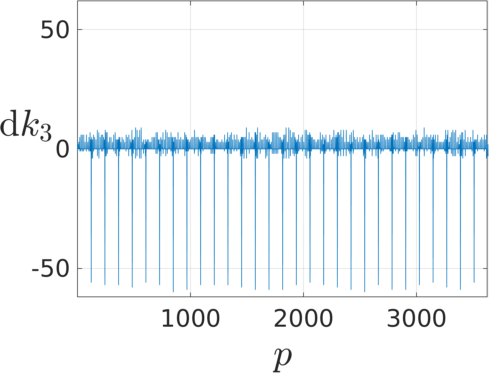}\end{overpic}\\\vspace{2mm}
\begin{overpic}[width=0.246\textwidth,draft=false]{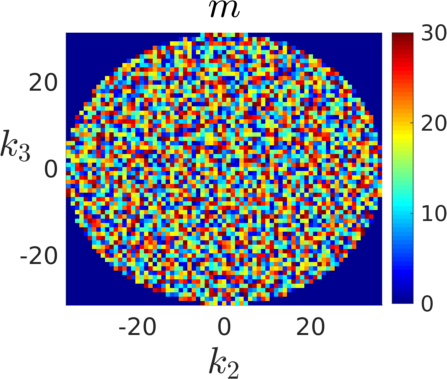}\put(0,0){\textbf{c)}}\end{overpic}
\begin{overpic}[width=0.246\textwidth,draft=false]{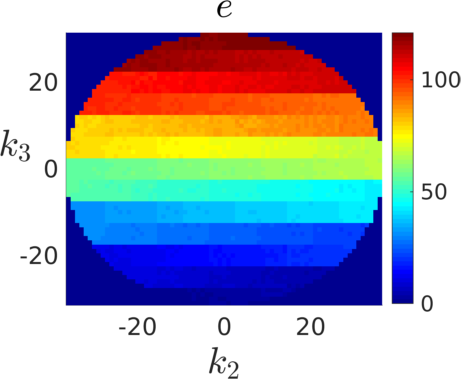}\end{overpic}
\begin{overpic}[width=0.246\textwidth,draft=false]{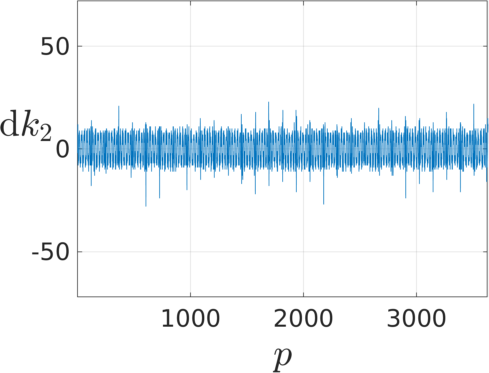}\end{overpic}
\begin{overpic}[width=0.246\textwidth,draft=false]{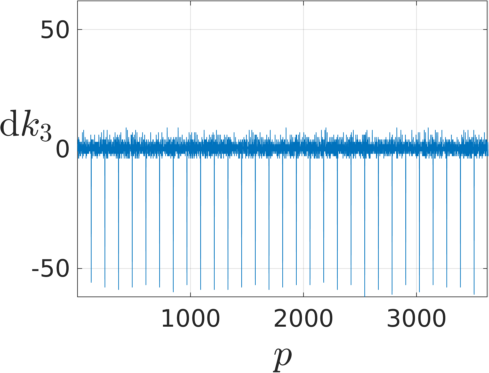}\end{overpic}
\end{center}\vspace{-5mm}
\input{caption3}
\end{figure}
\fi

\ifdefined\figures
\begin{figure}[!htb]
\begin{center}
\begin{overpic}[width=0.246\textwidth,draft=false]{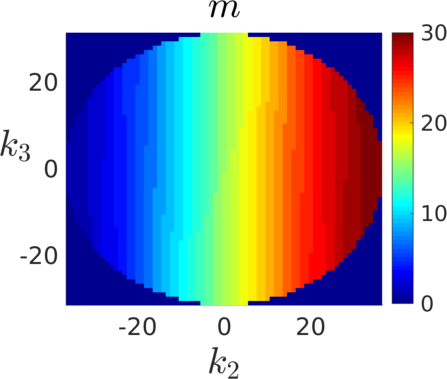}\put(0,0){\textbf{a)}}\end{overpic}
\begin{overpic}[width=0.246\textwidth,draft=false]{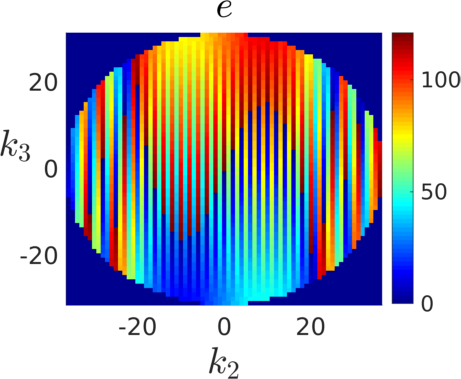}\end{overpic}
\begin{overpic}[width=0.246\textwidth,draft=false]{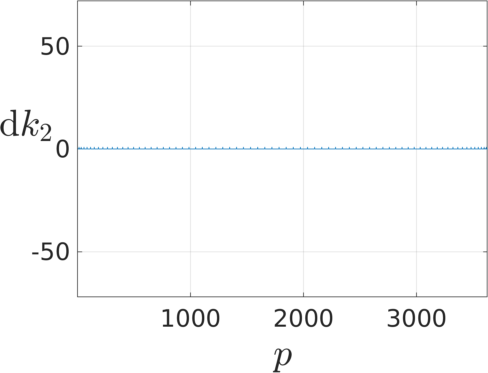}\end{overpic}
\begin{overpic}[width=0.246\textwidth,draft=false]{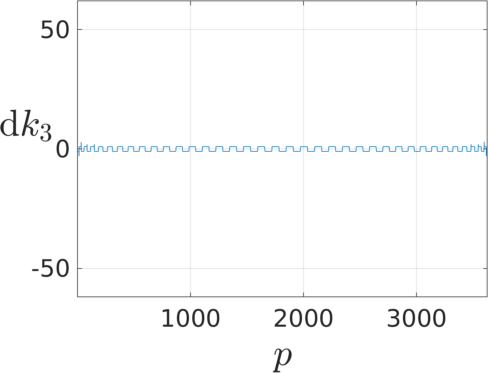}\end{overpic}\\\vspace{2mm}
\begin{overpic}[width=0.246\textwidth,draft=false]{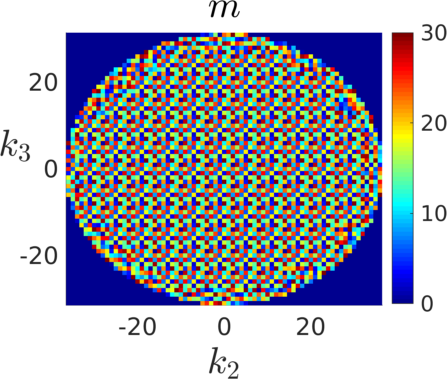}\put(0,0){\textbf{b)}}\end{overpic}
\begin{overpic}[width=0.246\textwidth,draft=false]{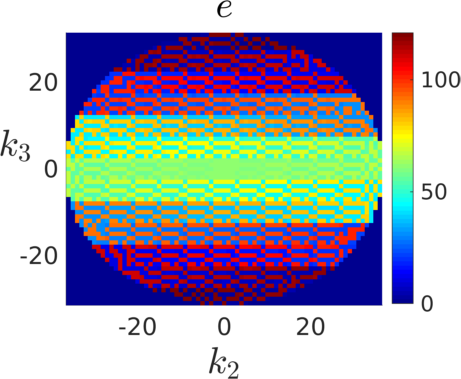}\end{overpic}
\begin{overpic}[width=0.246\textwidth,draft=false]{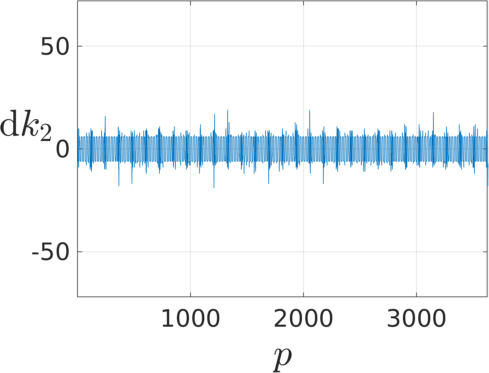}\end{overpic}
\begin{overpic}[width=0.246\textwidth,draft=false]{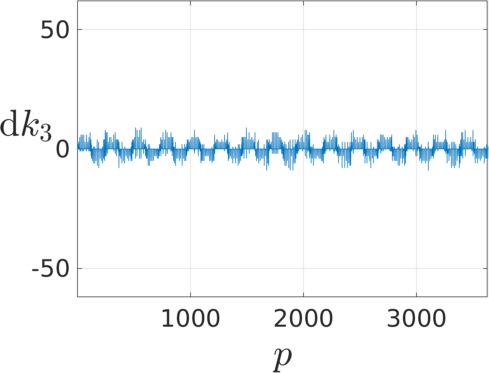}\end{overpic}\\\vspace{2mm}
\begin{overpic}[width=0.246\textwidth,draft=false]{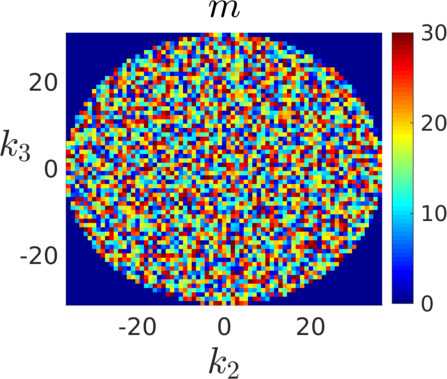}\put(0,0){\textbf{c)}}\end{overpic}
\begin{overpic}[width=0.246\textwidth,draft=false]{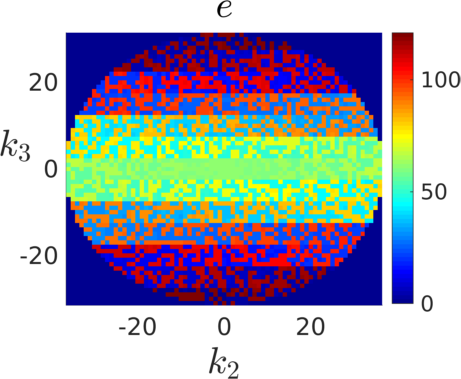}\end{overpic}
\begin{overpic}[width=0.246\textwidth,draft=false]{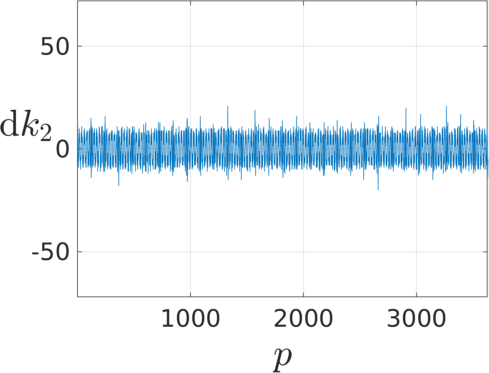}\end{overpic}
\begin{overpic}[width=0.246\textwidth,draft=false]{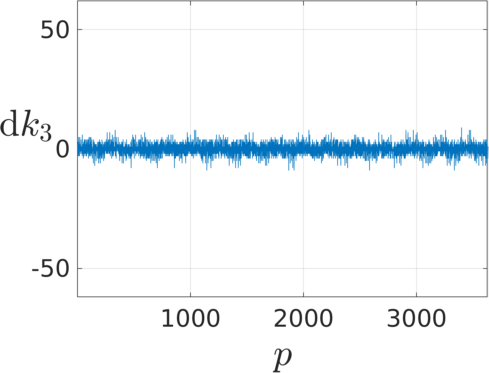}\end{overpic}\\\vspace{2mm}
\begin{overpic}[width=0.246\textwidth,draft=false]{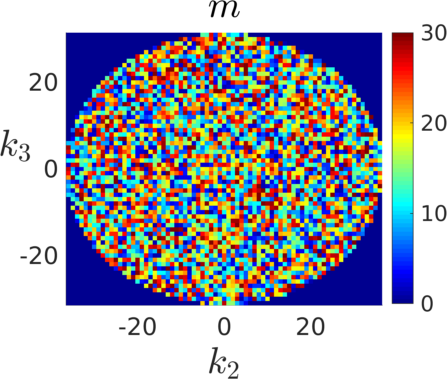}\put(0,0){\textbf{d)}}\end{overpic}
\begin{overpic}[width=0.246\textwidth,draft=false]{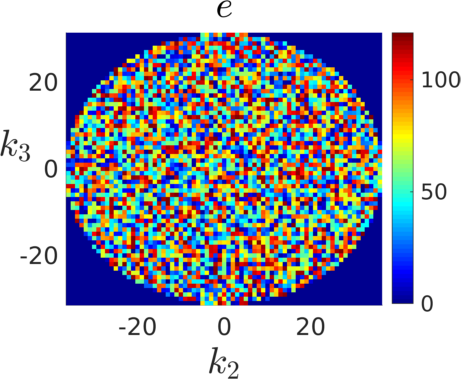}\end{overpic}
\begin{overpic}[width=0.246\textwidth,draft=false]{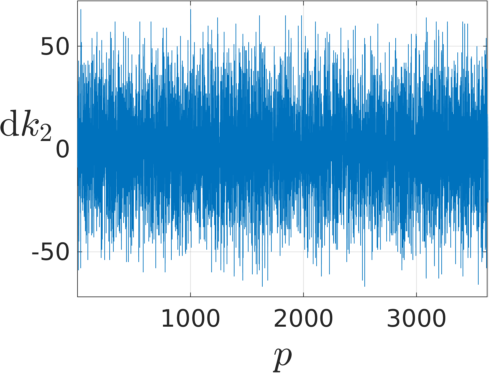}\end{overpic}
\begin{overpic}[width=0.246\textwidth,draft=false]{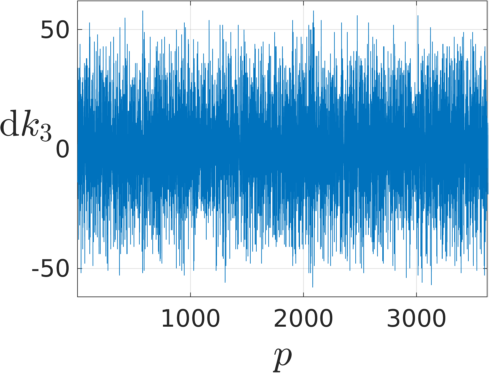}\end{overpic}
\end{center}\vspace{-5mm}
\input{caption4}
\end{figure}
\fi

\ifdefined\figures
\subsection{Aligned reconstruction refinements}
\else
\subsection*{Aligned reconstruction refinements}
\fi

\label{sec:REAS}

We propose a series of refinements for improved and more efficient aligned reconstructions:

\begin{itemize}
	\item \textbf{Spatial multiresolution}. The spatial and spectral grids for both subproblems in Eq.~\eqref{ec:GEFD} can be refined according to a given multiresolution pyramid as commonly used in image registration~\citep{Unser93}. In contrast to sequential sampling, the proposed orders allow for motion estimates from samples at coarse scales (area enclosed in cyan in Fig.~\ref{fig:SAOR}) to be completely exploited when reconstructing at fine scales. This is useful for quick aligned reconstructions as adequate motion estimates are often possible at coarse scales.
	\item \textbf{Temporal multiresolution}. Motion estimation can also be attempted at intra-shot (or intra-sweep) levels (for instance using the samples enclosed within the yellow areas for the $4$ intra-segment subdivisions in Fig.~\ref{fig:SAOR}). However, estimates using low spatial harmonics localize the structures at coarse scales only and, conversely, motion estimation using high spatial harmonics alone is limited by lower SNR and prone to local optima. These limitations can be alleviated using hierarchical estimation refinements by temporally subdividing the samples considered by the motion states within a shot in a coarse-to-fine manner.
	\item \textbf{Coil compression}. The two subproblems in Eq.~\eqref{ec:GEFD} can operate on a reduced number of virtual channels~\citep{Buehrer07}.
	\item \textbf{Motion compression}. The reconstruction subproblem complexity grows with the number of motion states, which can be reduced by motion compression or binning. Estimated motion parameter traces are approximated using piecewise constant functions by Haar wavelet decomposition truncation with threshold $\boldsymbol{\tau}$. Thereby, original motion states are packed into effective states by grouping together those contiguous states with similar motion parameters into an effective motion parameter vector $\tilde{\boldsymbol{\theta}}$. Thus, the reconstruction complexity is driven by the underlying motion complexity.
	\item \textbf{Robustness}. Accurate intra-shot corrections may be infeasible, for instance due to temporary inconsistencies in the magnetization. Denoting the real motion parameters by $\boldsymbol{\theta}^{\ast}$ we can ideally characterize the loss $r_{\mathbf{x},\boldsymbol{\theta}^{\ast}}$ using the sampling noise properties. Sampling noise follows a circularly symmetric complex Gaussian additive stationary distribution and, after whitening, it is independent across channels, so the losses per profile $r[m,e_m]=\sum_{c,k_1}r[m,e_m,c,k_1]$ should ideally follow a $\chi^2$ distribution. To account for the sensitivity of the residuals to the underlying signal, we use trimmed statistics on a logarithmic scale $r_b[m]=P^{\mathbb{E}_m}_{c(b)}\log(r[m,e_m])$ with $b\in{1,\ldots,B}$ indexing the $100c(b)\,\%$ centile $P_c$ of the loss distribution through k-space $\mathbb{E}_m$. As we are concerned with anomalously high residuals, robust estimates of the scale and mean of the statistic distribution across segments $\mathbb{M}$ are obtained respectively by $\sigma_b=\sqrt{2}(P^\mathbb{M}_{c_{\text{U}}}r_b[m]-P^{\mathbb{M}}_{c_{\text{L}}}r_b[m])/(\erfc^{-1}(2c_{\text{U}})-\erfc^{-1}(2c_{\text{L}}))$ and $\mu_b=P^{\mathbb{M}}_{(c_{\text{U}}+c_{\text{L}})/2}r_b[m]+\sqrt{2}\sigma_b\erfc^{-1}(c_{\text{U}}+c_{\text{L}})$ choosing $c_{\text{U}}=0.25$ and $c_{\text{L}}=0.125$. Using these estimates, the statistics are normalized and averaged into $\overline{r}[m]=\sum_b(r_b[m]-\mu_b)/B\sigma_b$, and segments are weighted in the reconstruction by a matrix $\mathbf{W}$ with entries $w[m]=\min(M\erfc(\overline{r}[m]/\sqrt{2})/(2\tau_w),1)$, with $\tau_w$ an acceptance threshold corrected for multiple comparisons.
	\item \textbf{Regularization}. If outlier segment rejection is activated or the reconstruction is applied to accelerated scans, some form of regularization may be advisable. This is considered by reformulating the reconstruction as:
\begin{equation}
\label{ec:RESH}
\hat{\mathbf{x}}^{(i+1)}=\displaystyle\argmin_{\mathbf{x}}\|\mathbf{W}^{1/2}(\mathbf{A}\boldsymbol{\mathcal{F}}\mathbf{S}\mathbf{T}_{\tilde{\boldsymbol{\theta}}^{(i)}}\mathbf{x}-\mathbf{y})\|_2^2+2\lambda\|\boldsymbol{\mathcal{S}}\mathbf{x}\|_1,
\end{equation}
where $\lambda$ controls the degree of regularization and $\boldsymbol{\mathcal{S}}$ corresponds to a shearlet decomposition, which provides nearly optimal approximation rates for piecewise smooth functions with discontinuities on a piecewise smooth surface~\citep{Kutyniok12}. We resort to an iteratively reweighted least squares (IRWLS) solver, able to produce high quality solutions in a few iterations \citep{Voronin17}, with $\lambda$ adaptively updated according to~\cite{Li18} using a normalized Rayleigh-quotient trace estimator~\citep{Avron11}.
\end{itemize}

%% file: caption1.tex
\caption{Sketch of volumetric sampling for an exemplary $K=4\times 4\times 4$ space (in radians). \textbf{a)} Measured samples after acquiring the first readout. \textbf{b)} Measured samples after acquiring the first segment. \textbf{c)} Measured samples after whole sequence acquisition with color coding used to differentiate each segment. \textbf{d)} View of the $k_2$-$k_3$ PE plane.}
\label{fig:VOEN}

%% file: caption2.tex
\caption{Example of segments for different encoding orders to cover a $P=K_2K_3$ PE plane (in radians) with $K_2=K_3=32$ using an acquisition partitioned into $M=16$ segments with equal number of profiles per segment $E_s=E=P/M=64$. Full set of profiles to cover the PE plane as red circles, profiles within a segment filled in blue, underlying tilings ($U_2=U_3=4$) in purple, samples at half the spatial resolution enclosed in cyan, and areas covered by $4$ intra-segment temporal subdivisions of considered segment enclosed in yellow. \textbf{a)} {\ber Sequential}; \textbf{b)} {\ber Checkered}; \textbf{c)} {\ber Random-checkered}; and \textbf{d)} {\ber Random} traversals.}
\label{fig:SAOR}

%% file: caption3.tex
\caption{Segment $m$ and tiling $e$ temporal orders together with trajectory derivatives $\text{d}k_2$ and $\text{d}k_3$ (left to right) for {\ber Zig-zag} tiling orders used in non steady-state sequences. The example corresponds to a case with $P=3630$ profiles, $M=30$ segments and tiling pattern $U_2\times U_3=6\times 5$. \textbf{a)} {\ber Sequential}; \textbf{b)} {\ber Checkered}; and \textbf{c)} {\ber Random-checkered} segments.}
\label{fig:ECORA}

%% file: caption4.tex
\caption{Segment $m$ and tiling $e$ temporal orders together with trajectory derivatives $\text{d}k_2$ and $\text{d}k_3$ (left to right) for {\ber Alternating zig-zag} tiling orders used in steady-state sequences. The example corresponds to a case with $P=3630$ profiles, $M=30$ segments and tiling pattern $U_2\times U_3=6\times 5$. \textbf{a)} {\ber Sequential}; \textbf{b)} {\ber Checkered}; \textbf{c)} {\ber Random-checkered}; and \textbf{d)} {\ber Random} segments.}
\label{fig:ECORB}

%% file: 3methods.tex
\ifdefined\figures
\section{Methods}
\else
\section*{Methods}
\fi

\label{sec:METH}

\ifdefined\figures
\subsection{Synthetic experiments}
\else
\subsection*{Synthetic experiments}
\fi

\label{sec:VADE}

Our contributions are validated using a synthetic dataset built from a $T_2$ neonatal brain axial ground truth (GT) image without perceptible motion artifacts. This corresponds to a multi-slice TSE sequence acquired on a $3\,\mbox{T}$ \textsmaller{\textsc{Philips Achieva TX}} (same scanner as for in-vivo tests) using a $C=32$-element neonatal head coil array, $0.8\times 0.8\,\mbox{mm}$ in-plane resolution, $1.6\,\mbox{mm}$ slice thickness, echo time $T_{\text{E}}=145\,\mbox{ms}$, repetition time $T_{\text{R}}=12\,\mbox{s}$, and flip angle $\alpha=90^{\circ}$. Coil sensitivities were estimated from a separate reference scan~\citep{Allison13}. We use a 2D dataset and no regularization or outlier rejection for a concise presentation of results. We assume that the simulated 2D k-space corresponds to the $k_2$-$k_3$ PE plane of 3D scans and expect the driving conclusions to be extensible to 3D because estimates should be easier along the missing fully-sampled readout direction.

Simulations were conducted to compare the conventional sequential order to the various proposed schemes as well as to characterize their performance. The forward model in the presence of rigid motion is applied to the GT to generate synthetically motion corrupted data. Synthesized measures are corrupted with noise levels corresponding to a mean SNR of $30\,$dB for reconstructions in the absence of motion or acceleration. Different degrees of motion are generated by drawing independent motion states uniformly at random on an interval of rotations $[-\theta/2,\theta/2]$ around the field of view (FOV) center. Satisfactory convergence in the presence of noise can be ascertained on the assumption of an identifiable global optimum basin by $r_{\hat{\mathbf{x}},\hat{\boldsymbol{\theta}}}\leq r_{\hat{\mathbf{x}},\boldsymbol{\theta}^{\ast}}$. In this case, the error in the motion parameters $\hat{\boldsymbol{\theta}}-\boldsymbol{\theta}^{\ast}$ is attributed to the uncertainty from the measurement noise and not to partial convergence. Note that we can generally achieve a lower loss for the joint problem ($r_{\hat{\mathbf{x}},\hat{\boldsymbol{\theta}}}$) than with the knowledge of the motion parameters ($r_{\hat{\mathbf{x}},\boldsymbol{\theta}^{\ast}}$) due to the larger complexity of the former. Reconstructions are terminated when $r_{\hat{\mathbf{x}},\hat{\boldsymbol{\theta}}}\leq r_{\hat{\mathbf{x}},\boldsymbol{\theta}^{\ast}}$ and the abscissa scale of the convergence plots was chosen so that iterations have a direct translation into computational costs.

\ifdefined\figures
\subsection{In-vivo experiments}
\else
\subsection*{In-vivo experiments}
\fi

\label{sec:MATE}

In-vivo experiments include the main families of volumetric sequences for brain MRI (see Table~\ref{tab:SEQU}). We have performed a controlled motion experiment on a consented adult volunteer and applied the method to replace sedation on pediatric subjects scanned after written informed parental consent for an epilepsy study. Imaging is performed using a $32$-channel adult head coil. Data is acquired in the inferior-superior (IS) $k_1$, anterior-posterior (AP) $k_2$ and left-right (LR) $k_3$ orientation using our scanner implementation of the {\ber Random-checkered} traversal. This way, potentially strongest rotations on the sagittal plane are captured by the $k_1$-$k_2$ coordinates, which may increase the resolvability of intra-shot motion. In addition, IS readouts allow to easily downweight additional motion sources within the FOV when estimating for motion by restricting the loss function to the superior part of the FOV ($2/3$ in our implementation). Finally, this orientation facilitates non-selective RF excitation pulses for shorter $T_{\text{R}}$.
\ifdefined\figures
\input{table1}
\fi

In the controlled motion experiment, the volunteer was first scanned without deliberate motion, and then asked to perform extreme and continuous motion for the entire scan, which was repeated three times. Reconstructed volumes are jointly registered together for error comparisons. The pediatric cohort includes $26$ subjects ranging from $3$ to $19$ years old (mean$\pm$std of $12\pm 5$ years), typically acquiring one MP-RAGE, TSE and FLAIR, two SPGRs and three bSSFPs for an approximate total of $208$ tested volumes across all participants. Strongest artifacts in our data are generally arising from motion, so the reported case has been separately chosen for each modality as the most artifacted after reconstruction without motion correction.

\ifdefined\figures
\subsection{Implementation details}
\else
\subsection*{Implementation details}
\fi

\label{sec:IMDE}

In the in-vivo experiments sensitivities are compressed into a number of channels corresponding to a $10\%$ SNR loss. The number of resolution levels is defined as $L=\displaystyle\left\lfloor\log_2(4\,\mbox{mm}/\Delta_{\mathbf{y}}^{\text{min}})\right\rfloor+1$, with $\lfloor\cdot\rfloor$ denoting the biggest integer lower or equal than the argument and $\Delta_{\mathbf{y}}^{\text{min}}$ the minimum of the voxel sizes along different directions. As we use $2\times$ subsampling ratios, we operate at a minimum resolution of $4\,\mbox{mm}$. In the first iteration at level $l$, a soft-masked~\citep{Fuderer04} full CG reconstruction is run till loss reduction saturation. Then, the method quickly alternates between reconstruction and motion correction using one CG and one LM iteration with heuristically updated damping and line search. We activate a flag for provisional convergence of the parameters of a given motion state when the maximum update is smaller than a threshold $\boldsymbol{\tau}_{\Delta\boldsymbol{\theta}}=\{0.05,0.02^{\circ}\mbox{mm}^{-1}\}\Delta_{\mathbf{y}^l}^{\text{max}}$, with same values used for motion compression. This saves computations by considering motion updates only on not-converged parameters. However, this flag is reset to $0$ whenever $i=n(n-1)/2+1$ ($n\in\mathbb{N}_{>0}$) to account for the impact of the updated reconstructions in the motion parameter estimates. Joint convergence is achieved when provisional convergence is achieved for all motion states. Then, the method runs a full CG reconstruction with the consolidated motion parameters. If regularized outlier rejection reconstructions are activated, artifacted segments are rejected at levels $l$ such that $\Delta_{\mathbf{y}^l}^{\text{max}}\leq 2\,\mbox{mm}$ by densely sampling within $[c(1),c(B)]=0.5+(0.35/\max(\Delta_{\mathbf{y}^l}^{\text{max}},1))[-1,1]$ and using $\tau_w=0.05$. If regularization is applied, shearlets are designed based on~\cite{Kutyniok16} and a final reconstruction is launched with $3$ CG iterations within $2$ updates of the IRWLS-induced cost function and $\lambda$. Reconstructions are performed on a $8(16)\times$ \textsmaller{\textsc{Intel(R) Core(TM) i7-5960X}} $3.00\,\mbox{GHz}$ CPU, $64\,\mbox{GB}$ RAM, \textsmaller{\textsc{GeForce GTX TITAN X}} GPU. For further implementation details, readers can refer to the source code. 

%% file: table1.tex
\begin{table}[!htb]
\begin{tiny}
\begin{center}
\scalebox{\ifdefined\figures 1.05\else 1.1\fi}{
\begin{tabular}{c|c|c|c|c|c|c|c|c|c|c|c}
\hline
Subjects & Modality & $T_{\text{R}}$ & $T_{\text{E}}$ & $T_{\text{I}}$ & $\alpha$ & $P$ & $U_2\times U_3$ & $\Delta_{\mathbf{y}}$ & $R$ & $\text{FOV}_{\mathbf{x}}$ & $t$ \\\hline\hline
Adult & MP-RAGE & $8.3\,\mbox{ms}$ & $3.7\,\mbox{ms}$ & $1.00\,\mbox{s}$ & $8.0^{\circ}$ & $3630$ & $6\times 5$ & $1.5\,\mbox{mm}$ & $2\times 2$ & $240\times 240\times 220\,\mbox{mm}$ & $1\,\mbox{min}\,29\,\mbox{s}$ \\\hline
\multirow{5}{*}{Pediatric} & MP-RAGE & $7.0\,\mbox{ms}$ & $2.3\,\mbox{ms}$ & $0.90\,\mbox{s}$ & $8.0^{\circ}$ & $17920$ & $8\times 10$ & \multirow{5}{*}{$1.0\,\mbox{mm}$} & \multirow{3}{*}{$1.4\times 1.4$} & \multirow{5}{*}{$240\times 240\times 188\,\mbox{mm}$} & $3\,\mbox{min}\,07\,\mbox{s}$ \\\cline{2-8}\cline{12-12}
& TSE & $2.50\,\mbox{s}$ & $0.31\,\mbox{s}$ & -- & \multirow{2}{*}{$90.0^{\circ}$} & $17955$ & $9\times 15$ &  &  &  & $5\,\mbox{min}\,42\,\mbox{s}$ \\\cline{2-5}\cline{7-8}\cline{12-12}
& FLAIR & $5.00\,\mbox{s}$ & $0.41\,\mbox{s}$ & $1.80\,\mbox{s}$ &  & $18200$ & \multirow{3}{*}{$10\times 10$} &  &  &  & $8\,\mbox{min}\,30\,\mbox{s}$ \\\cline{2-7}\cline{10-10}\cline{12-12}
& SPGR & $8.2\,\mbox{ms}$ & $3.0\,\mbox{ms}$ & \multirow{2}{*}{--} & $12.3^{\circ}$ & \multirow{2}{*}{$21654$} &  &  & \multirow{2}{*}{$1.3\times 1.3$} &  & $3\,\mbox{min}\,06\,\mbox{s}$ \\\cline{2-4}\cline{6-6}\cline{12-12}
& bSSFP & $6.0\,\mbox{ms}$ & $3.0\,\mbox{ms}$ &  & $46.0^{\circ}$ &  &  &  &  &  & $2\,\mbox{min}\,16\,\mbox{s}$ \\\hline
\end{tabular}
}
\end{center}
\end{tiny}
\ifdefined\figures
\ifdefined\comments
\else
\vspace{-4mm}
\fi
\fi
\caption{Sequence parameters for the different modalities considered in the experiments. $T_{\text{I}}$ is the inversion time, $\Delta_{\mathbf{y}}$ the acquired resolution, $R$ the uniform acceleration factor, $\text{FOV}_{\mathbf{x}}$ the reconstructed FOV, and $t$ the scan duration.}
\label{tab:SEQU}
\end{table}

%% file: 4results.tex
\ifdefined\figures
\section{Results}
\else
\section*{Results}
\fi

\label{sec:RESU}

\ifdefined\figures
\subsection{Validation}
\else
\subsection*{Validation}
\fi

In Fig.~\ref{fig:VACO} we compare different simulated reconstruction scenarios showing the losses when iterating the method, $r_{\hat{\mathbf{x}}^{(i+1)},\hat{\boldsymbol{\theta}}^{(i)}}$ as solid colored lines with joint iterations represented by markers. Losses in the convergence plots are normalized to the minimum of the reference levels $r_{\hat{\mathbf{x}},\boldsymbol{\theta}^{\ast}}$, which are shown as dashed lines strongly overlapped for the different alternatives. Fig.~\ref{fig:VARE} includes reconstructions with and without motion correction for different reconstruction scenarios and provides absolute value error maps and mean SNR for the compared cases.

\ifdefined\figures
\begin{figure}[!htb]
\begin{center}
\begin{overpic}[width=0.99\textwidth,draft=false]{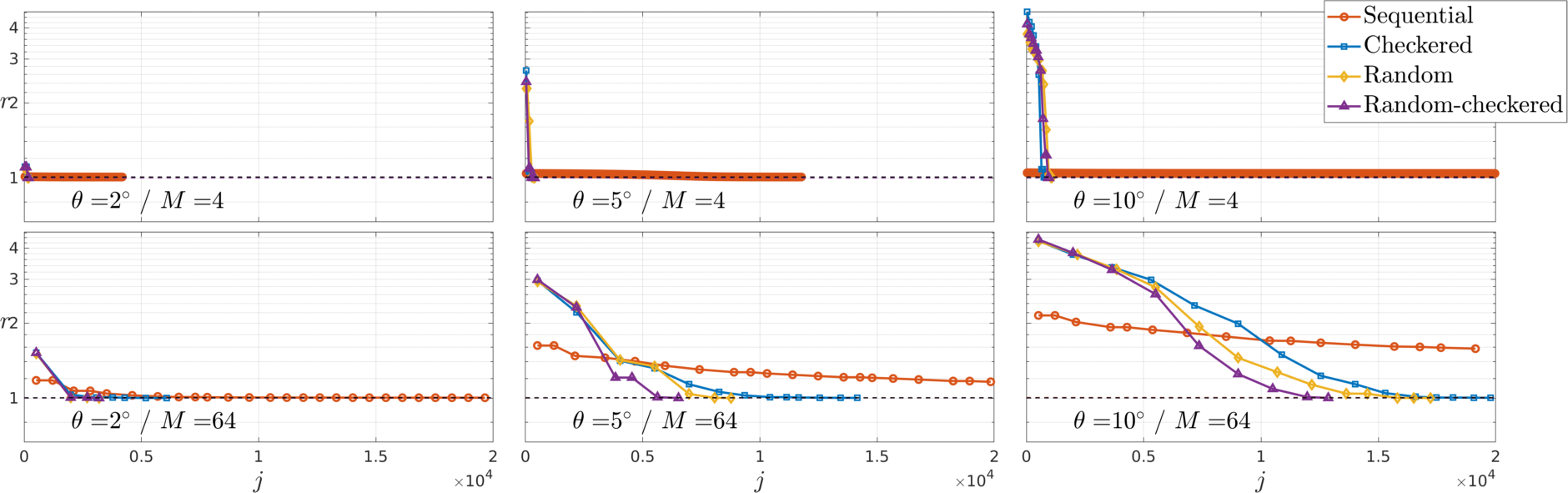}\put(0,0){\textbf{a)}}\end{overpic}\vspace{2mm}
\begin{overpic}[width=0.99\textwidth,draft=false]{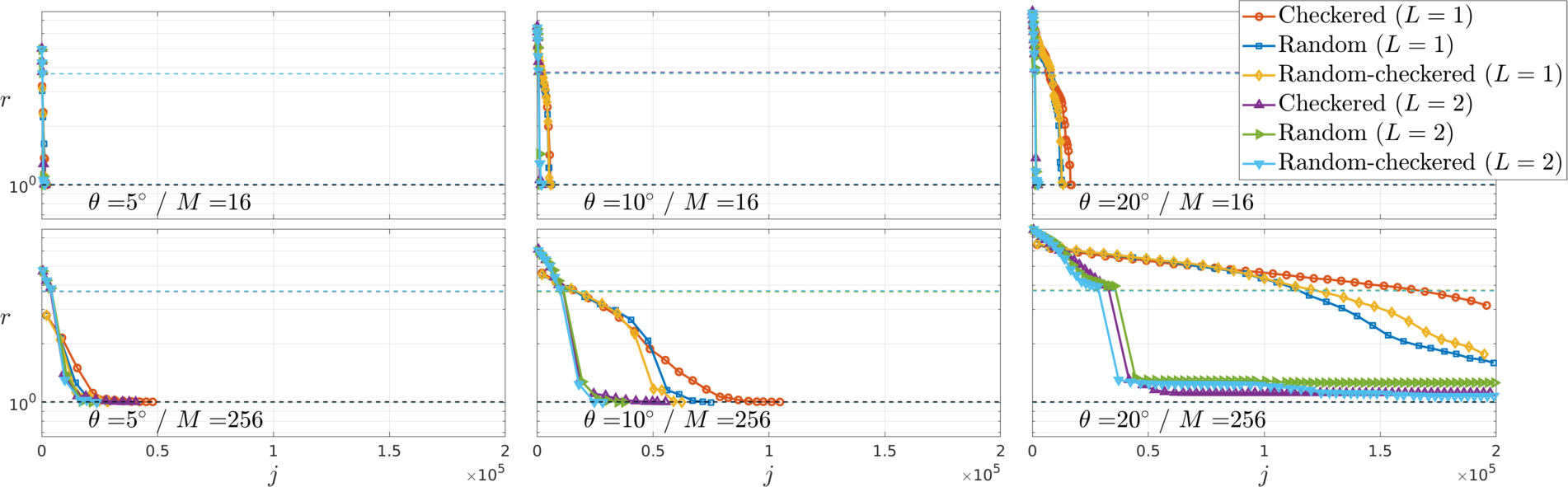}\put(0,0){\textbf{b)}}\end{overpic}\vspace{2mm}
\begin{overpic}[width=0.99\textwidth,draft=false]{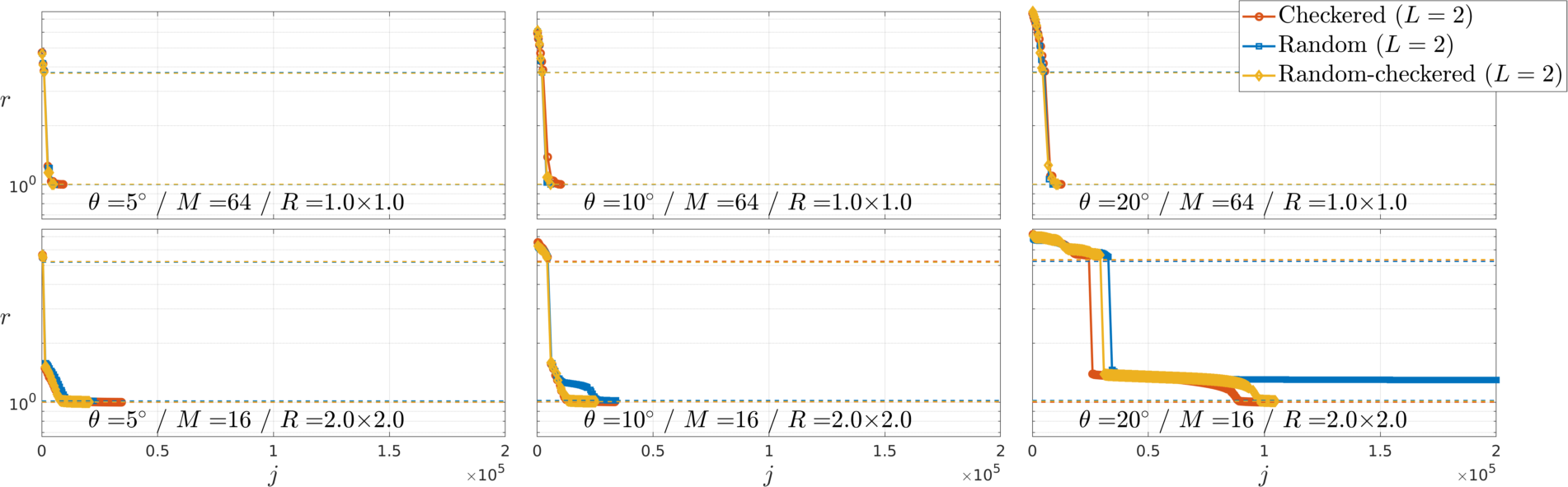}\put(0,0){\textbf{c)}}\end{overpic}
\end{center}
\input{caption5}
\end{figure}
\fi

\ifdefined\figures
\begin{figure}[!htb]
\begin{overpic}[width=0.997\textwidth,draft=false]{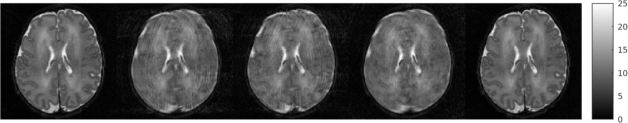}\end{overpic}\vspace{2mm}\\
\begin{overpic}[width=\textwidth,draft=false]{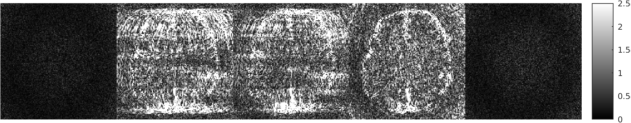}\put(0,-4){\textbf{a)}}\footnotesize{
\put(9.2593,-4){\makebox[0pt]{\Centerstack{No motion\\$\mbox{SNR}=29.99\,\mbox{dB}$}}}
\put(27.7778,-4){\makebox[0pt]{\Centerstack{Uncorrected\\{\ber Sequential}\\$\mbox{SNR}=12.72\,\mbox{dB}$}}}
\put(46.2963,-4){\makebox[0pt]{\Centerstack{Corrected\\{\ber Sequential}\\$\mbox{SNR}=13.37\,\mbox{dB}$}}}
\put(64.8148,-4){\makebox[0pt]{\Centerstack{Uncorrected\\{\ber Random-checkered}\\$\mbox{SNR}=14.51\,\mbox{dB}$}}}
\put(83.3333,-4){\makebox[0pt]{\Centerstack{Corrected\\{\ber Random-checkered}\\$\mbox{SNR}=28.40\,\mbox{dB}$}}}
}\end{overpic}\vspace{14mm}
\begin{overpic}[width=0.997\textwidth,draft=false]{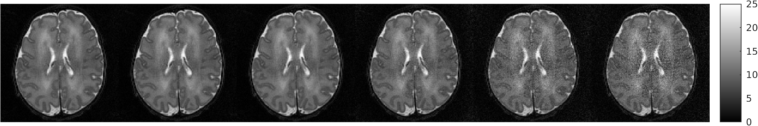}\end{overpic}\vspace{2mm}\\
\begin{overpic}[width=\textwidth,draft=false]{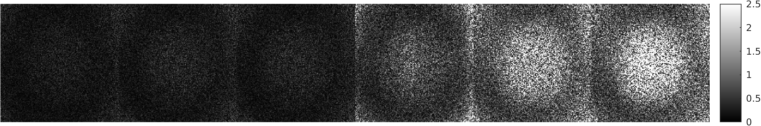}\put(0,-4){\textbf{b)}}\footnotesize{
\put(7.8182,-4){\makebox[0pt]{\Centerstack{$R=1\times 1$\\No motion\\$\mbox{SNR}=29.98\,\mbox{dB}$}}}
\put(23.4545,-4){\makebox[0pt]{\Centerstack{$R=1\times 1$\\Known motion\\$\mbox{SNR}=28.42\,\mbox{dB}$}}}
\put(39.0909,-4){\makebox[0pt]{\Centerstack{$R=1\times 1$\\Estimated motion\\$\mbox{SNR}=28.40\,\mbox{dB}$}}}
\put(54.7273,-4){\makebox[0pt]{\Centerstack{$R=2\times 2$\\No motion\\$\mbox{SNR}=20.78\,\mbox{dB}$}}}
\put(70.3636,-4){\makebox[0pt]{\Centerstack{$R=2\times 2$\\Known motion\\$\mbox{SNR}=16.42\,\mbox{dB}$}}}
\put(86,-4){\makebox[0pt]{\Centerstack{$R=2\times 2$\\Estimated motion\\$\mbox{SNR}=15.02\,\mbox{dB}$}}}
}\end{overpic}\vspace{10mm}
\input{caption6}
\end{figure}
\fi

\ifdefined\figures
\subsubsection{Encoding orders}
\else
\subsubsection*{Encoding orders}
\fi

Fig.~\ref{fig:VACO}a compares the {\ber Sequential}, {\ber Checkered}, {\ber Random-checkered} and {\ber Random} traversals. Global convergence is achieved for all considered $M$ and $\theta$ when using any of the {\ber Checkered}, {\ber Random-checkered} or {\ber Random} traversals. In contrast, when using the {\ber Sequential} order, the method converges to a local optimum or fails to converge in the prescribed iterations except for $\theta\in\{2^{\circ},5^{\circ}\}$ / $M=4$. The loss at the first iteration $r_{\hat{\mathbf{x}}^{(1)},\hat{\boldsymbol{\theta}}^{(0)}}$ is always bigger when using non-sequential traversals. This increased inconsistency in the measurement domain relates to the aligned reconstruction sensitivity to motion degradation.

Fig.~\ref{fig:VARE}a shows reconstructions and errors with and without motion correction for the {\ber Sequential} and {\ber Random-checkered} traversals together with GT motion-free reconstructions. Motion corrected reconstructions using the {\ber Random-checkered} data appear similar to the GT despite the strong blurring in uncorrected reconstructions. This is confirmed by the lack of perceptible structure in the residuals and a moderate noise amplification. In contrast, corrections using the {\ber Sequential} traversal provide only a modest visual benefit.

\ifdefined\figures
\subsubsection{Multiresolution}
\else
\subsubsection*{Multiresolution}
\fi

Fig.~\ref{fig:VACO}b compares the {\ber Checkered}, {\ber Random-checkered} and {\ber Random} traversals when using a single scale for joint motion estimation and reconstruction ($L=1$) and when first approximating the motion solution at half the acquired resolution to initialize the joint problem at full resolution ($L=2$). The {\ber Sequential} traversal was excluded because, as discussed when introducing the multiresolution strategy, it has no opportunity to improve from the poor relative performance showed in Fig.~\ref{fig:VACO}a by exploiting multiresolution. Plots also include $r_{\hat{\mathbf{x}},\boldsymbol{\theta}^{\ast}}$ at the coarse scale. Global convergence is achieved for all traversals at all considered configurations except at $M=256$ / $\theta=20^{\circ}$. However, the multiresolution strategy ($L=2$) achieves global convergence in less iterations or provides a solution with lower residuals ($M=256$ / $\theta=20^{\circ}$). For moderate levels of motion, convergence is generally quick. For instance, it takes approximately $10$ joint iterations $i$ when using the {\ber Random-checkered} traversal in a case where random excursions of up to $\theta=10^{\circ}$ are imposed in every one of the $M=256$ segments, probably a more challenging scenario than expected in practice.

\ifdefined\figures
\subsubsection{Acceleration}
\else
\subsubsection*{Acceleration}
\fi

Fig.~\ref{fig:VACO}c tests the ability of the {\ber Checkered}, {\ber Random-checkered} and {\ber Random} traversals (using $L=2$ scales) to operate in uniformly accelerated regimes as given by different acceleration factors $R$. We observe convergence to the global solution in all tested scenarios apart from the {\ber Random} traversal at $R=2\times 2$ / $M=16$ / $\theta=20^{\circ}$. Considering all conducted simulations, the {\ber Random-checkered} traversal is generally providing the quickest solutions.

Fig.~\ref{fig:VARE}b shows an example of reconstructions and errors in the absence of motion, with known motion and with estimated motion at $R=1\times 1$ and $R=2\times 2$. The SNR figures for $R=1\times 1$ and known motion show a degradation of $1.56\,\text{dB}$ with respect to the reference due to noise amplification from non-uniform effective k-space sampling after motion. No further degradation is introduced from motion estimation errors, as approximately the same SNR figures are obtained for known and estimated motion. $R=2\times 2$ acceleration in the absence of motion introduces a degradation of $9.20\,\text{dB}$ with respect to the reference, which stems from the reduced number of samples and the g-factor~\citep{Pruessmann99}. The presence of motion adds further degradations quantified as $4.36\,\text{dB}$, thus stronger than in the non-accelerated case. Therefore, without regularization, the limiting reconstruction quality in the presence of motion decreases with larger distances between neighboring k-space points. Finally, accelerating the scan has also an impact in the uncertainty of motion estimates, as we observe a degradation of $1.40\,\text{dB}$ from known to estimated motion, although the errors show no perceptible structure.

\ifdefined\figures
\subsection{Redundancy for motion tolerance}
\else
\subsection*{Redundancy for motion tolerance}
\fi

\label{sec:RERE}

Fig.~\ref{fig:READ} compares reconstructions without correction, with inter-shot corrections and when activating intra-shot corrections in the presence of extreme motion during in-vivo data acquisition. Intra-shot corrections are triggered by subsequent temporal binary subdivisions of the sampled information within each shot until $16$ motion states are estimated per shot. We show reconstructions without deliberate motion (GT reconstructions), and reconstructions and absolute differences with respect to the GT using $Q=1$, $Q=2$ and $Q=3$ repeats under extreme motion (Extreme motion reconstructions / errors, $Q=\{1,2,3\}$). Results for $Q=1$ and $Q=2$ correspond to the first repeats, with no remarkable differences observed when choosing any other combination. Reconstructions are provided without regularization or outlier rejection. Results without deliberate motion show that inter- and intra-shot corrections do not reduce the reconstruction quality, which demonstrates a safe application of generalized reconstructions in the absence of motion. Degradation is noticeable for uncorrected reconstructions in the presence of motion for all values of $Q$, although with less coherent ghosts as $Q$ increases due to incoherent blurring by {\ber Random-checkered} motion averaging. Inter-shot corrections increase the reconstruction quality in all cases, with more finely resolved cortical structures as $Q$ increases but with noticeably inferior quality than without deliberate motion. Residual degradation is only partially accounted when using intra-shot corrections on a single repeat, but can be more satisfactorily addressed with $Q=2$ and even more with $Q=3$. Namely, the level of deblurring in the fourth and sixth columns of Fig.~\ref{fig:READ}c makes corresponding reconstructions visually comparable to those of the first column despite the extreme and continuous motion (estimated excursions up to $25^{\circ}$). Thus, we can reason that powerful tolerance is achieved for $R=2\times 2$ and $Q=2$, so that $Q=1$ with acceleration $R=\sqrt{2}\times\sqrt{2}$\footnote{Both alternatives involve the same scanning time but the latter generates a lower g-factor. The former was used in this experiment because it was more convenient to our scanner implementation of the traversals.} may be adequate for motion tolerance in practice, which has been used to guide the acceleration in the pediatric cohort (see Table~\ref{tab:SEQU}). However, in contrast to computation times of $2\,\mbox{min}$ (non-deliberate motion), and $11\,\mbox{min}$ (extreme motion, $Q=3$) for inter-shot corrections, corresponding intra-shot corrections required $21\,\mbox{min}$ and $20\,\mbox{h}\,36\,\mbox{min}$. Thus, despite being technically feasible, intra-shot corrections may have limited applicability due to computational costs. Computational cost increase with the complexity of motion is due to the larger number of iterations for convergence and to the proposed motion compression strategy, with $13/30$ binned inter-shot motion states without deliberate motion and $30/30$ with extreme motion ($Q=1$), with proportional savings in the reconstruction steps in the former.
\ifdefined\figures
\begin{figure}[!htb]
\begin{center}
\begin{overpic}[width=0.138\textwidth,draft=false]{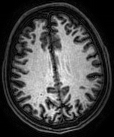}\put(1,2){\textcolor{white}{\textbf{a)}}}\end{overpic}
\begin{overpic}[width=0.138\textwidth,draft=false]{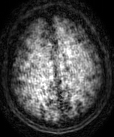}\end{overpic}
\begin{overpic}[width=0.138\textwidth,draft=false]{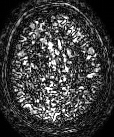}\end{overpic}
\begin{overpic}[width=0.138\textwidth,draft=false]{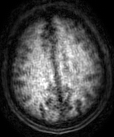}\end{overpic}
\begin{overpic}[width=0.138\textwidth,draft=false]{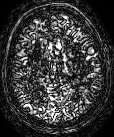}\end{overpic}
\begin{overpic}[width=0.138\textwidth,draft=false]{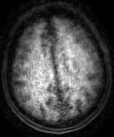}\end{overpic}
\begin{overpic}[width=0.138\textwidth,draft=false]{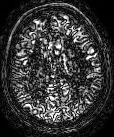}\end{overpic}\\
\begin{overpic}[width=0.138\textwidth,draft=false]{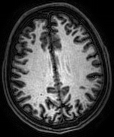}\put(1,2){\textcolor{white}{\textbf{b)}}}\end{overpic}
\begin{overpic}[width=0.138\textwidth,draft=false]{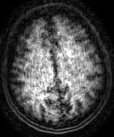}\end{overpic}
\begin{overpic}[width=0.138\textwidth,draft=false]{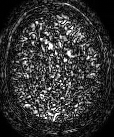}\end{overpic}
\begin{overpic}[width=0.138\textwidth,draft=false]{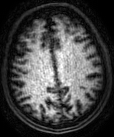}\end{overpic}
\begin{overpic}[width=0.138\textwidth,draft=false]{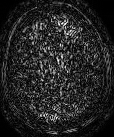}\end{overpic}
\begin{overpic}[width=0.138\textwidth,draft=false]{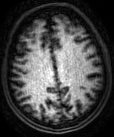}\end{overpic}
\begin{overpic}[width=0.138\textwidth,draft=false]{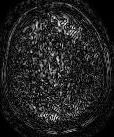}\end{overpic}\\
\begin{overpic}[width=0.138\textwidth,draft=false]{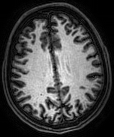}\put(1,2){\textcolor{white}{\textbf{c)}}}\footnotesize{\put(42.85,-26){\makebox[0pt]{\Centerstack{GT\\reconstructions}}}}\end{overpic}
\begin{overpic}[width=0.138\textwidth,draft=false]{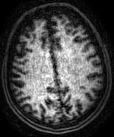}\footnotesize{\put(42.85,-26){\makebox[0pt]{\Centerstack{Extreme motion\\reconstructions\\$Q=1$}}}}\end{overpic}
\begin{overpic}[width=0.138\textwidth,draft=false]{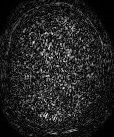}\footnotesize{\put(42.85,-26){\makebox[0pt]{\Centerstack{Extreme motion\\errors\\$Q=1$}}}}\end{overpic}
\begin{overpic}[width=0.138\textwidth,draft=false]{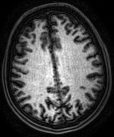}\footnotesize{\put(42.85,-26){\makebox[0pt]{\Centerstack{Extreme motion\\reconstructions\\$Q=2$}}}}\end{overpic}
\begin{overpic}[width=0.138\textwidth,draft=false]{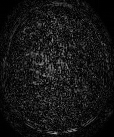}\footnotesize{\put(42.85,-26){\makebox[0pt]{\Centerstack{Extreme motion\\errors\\$Q=2$}}}}\end{overpic}
\begin{overpic}[width=0.138\textwidth,draft=false]{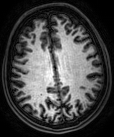}\footnotesize{\put(42.85,-26){\makebox[0pt]{\Centerstack{Extreme motion\\reconstructions\\$Q=3$}}}}\end{overpic}
\begin{overpic}[width=0.138\textwidth,draft=false]{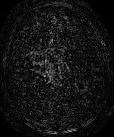}\footnotesize{\put(42.85,-26){\makebox[0pt]{\Centerstack{Extreme motion\\errors\\$Q=3$}}}}\end{overpic}
\\\vspace{6mm}
\end{center}
\input{caption7}
\end{figure}
\fi

\ifdefined\figures
\subsection{Non-compliant subjects}
\else
\subsection*{Non-compliant subjects}
\fi

\label{sec:INVI}

Fig.~\ref{fig:REIN} shows worst-case reconstructions without motion correction, with motion correction alone and with motion correction and the regularized outlier segment rejection. Results are shown for main structural brain MRI modalities, MP-RAGE, TSE, FLAIR, SPGR and bSSFP. In all sequences we observe a substantial improvement when activating motion-corrected reconstructions alone, with better delineated cortical structures. However, subtle artifacts are still present, either in the form of ghosts or of coloured noise. Fig.~\ref{fig:REIN}c shows that quality can be further improved by rejecting the less consistent segments and performing a regularized reconstruction. In some sequences discarding the artifacted segments seems to reduce residual artifacts from uncorrected fast motion (see for instance fine details in SPGR) while in others it seems to mainly improve the magnetization consistency (see TSE contrast). Across the cohort we have observed that motion artifact levels always decrease when compensating for motion, with no remarkable differences when activating the corrections in the absence of artifacts. This is along the lines of the quantitative population metrics obtained for the less favourable sequential sampling~\citep{Cordero-Grande16} or for multi-slice scans~\citep{Cordero-Grande18}. Worst-case results of Fig.~\ref{fig:REIN} have been judged satisfactory by the practitioners and researchers involved in the project. Therefore, the proposed methodology is delivering reliable examinations for unsedated pediatric subjects challenging to comply to the MRI motion requirements. In this experiment motion estimates were performed at half the acquired resolution with joint motion estimation and reconstruction always taking less time than final reconstructions at full resolution. Computation times range between $5\,\mbox{min}$ in least artifacted and $40\,\mbox{min}$ in most artifacted volumes in our cohort. Estimated motion traces and outlier segments are reported in \textsl{Supporting Information} Fig.~S1.

\ifdefined\figures
\begin{figure}[!htb]
\begin{center}
\begin{overpic}[width=0.195\textwidth,draft=false]{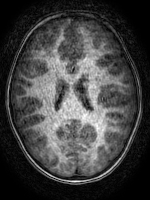}\put(1,2){\textcolor{white}{\textbf{a)}}}\end{overpic}
\begin{overpic}[width=0.195\textwidth,draft=false]{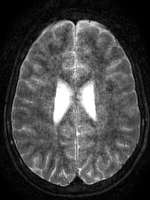}\end{overpic}
\begin{overpic}[width=0.195\textwidth,draft=false]{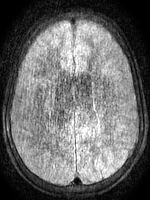}\end{overpic}
\begin{overpic}[width=0.195\textwidth,draft=false]{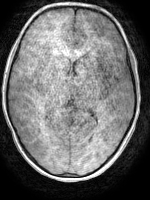}\end{overpic}
\begin{overpic}[width=0.195\textwidth,draft=false]{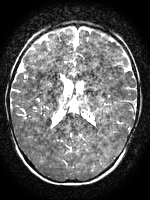}\end{overpic}\\
\begin{overpic}[width=0.195\textwidth,draft=false]{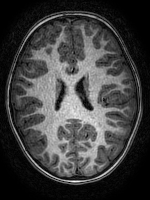}\put(1,2){\textcolor{white}{\textbf{b)}}}\end{overpic}
\begin{overpic}[width=0.195\textwidth,draft=false]{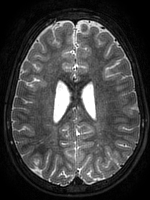}\end{overpic}
\begin{overpic}[width=0.195\textwidth,draft=false]{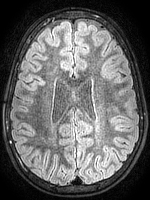}\end{overpic}
\begin{overpic}[width=0.195\textwidth,draft=false]{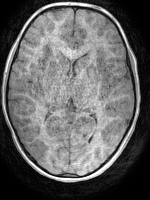}\end{overpic}
\begin{overpic}[width=0.195\textwidth,draft=false]{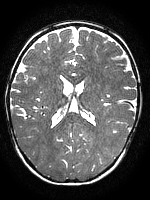}\end{overpic}\\
\begin{overpic}[width=0.195\textwidth,draft=false]{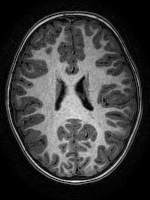}\put(1,2){\textcolor{white}{\textbf{c)}}}\footnotesize{\put(37.5,-9){\makebox[0pt]{\Centerstack{MP-RAGE}}}}\end{overpic}
\begin{overpic}[width=0.195\textwidth,draft=false]{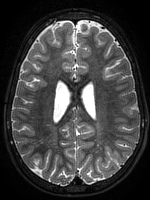}\footnotesize{\put(37.5,-9){\makebox[0pt]{\Centerstack{TSE}}}}\end{overpic}
\begin{overpic}[width=0.195\textwidth,draft=false]{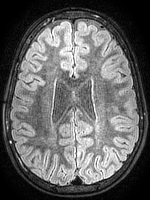}\footnotesize{\put(37.5,-9){\makebox[0pt]{\Centerstack{FLAIR}}}}\end{overpic}
\begin{overpic}[width=0.195\textwidth,draft=false]{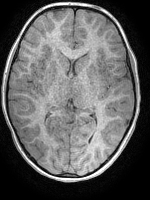}\footnotesize{\put(37.5,-9){\makebox[0pt]{\Centerstack{SPGR}}}}\end{overpic}
\begin{overpic}[width=0.195\textwidth,draft=false]{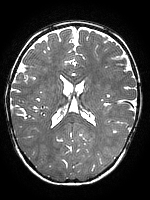}\footnotesize{\put(37.5,-9){\makebox[0pt]{\Centerstack{bSSFP}}}}\end{overpic}\\
\vspace{0mm}
\end{center}
\input{caption8}
\end{figure}
\fi

%% file: caption5.tex
\caption{Aligned reconstruction convergence against effective iterations $j$ defined as a single application of the encoding or decoding operator for a single motion state and coil channel at full resolution. \textbf{a)} Different encoding orders, number of segments $M\in\{4,64\}$ (rows), motion levels $\theta\in\{2^{\circ},5^{\circ},10^{\circ}\}$ (columns), $j_{\text{max}}=20000$. \textbf{b)} Different encoding orders and number of multiresolution levels $L$, number of segments $M\in\{16,256\}$ (rows), motion levels $\theta\in\{5^{\circ},10^{\circ},20^{\circ}\}$ (columns), $j_{\text{max}}=200000$. \textbf{c)} Different encoding orders, $L=2$, acceleration factors $R\in\{1\times 1,2\times 2\}$ with matched number of segments $M\in\{64,16\}$ (rows), motion levels $\theta\in\{5^{\circ},10^{\circ},20^{\circ}\}$ (columns), $j_{\text{max}}=200000$.}
\label{fig:VACO}

%% file: caption6.tex
\caption{Reconstruction results \textbf{a)} without and with motion correction for {\ber Sequential} and {\ber Random-checkered} traversals compared to the GT (top row) and corresponding error maps (bottom row) for $M=64$ / $\theta=10^{\circ}$ and \textbf{b)} for a {\ber Random-checkered} traversal without motion, and with known and estimated $\theta=10^{\circ}$ motion for $R=1\times 1$ and $R=2\times 2$ (top row) and corresponding error maps (bottom row).}
\label{fig:VARE}

%% file: caption7.tex
\caption{Reconstruction results for extreme motion in vivo. \textbf{a)} Uncorrected; \textbf{b)} inter-shot corrections; and \textbf{c)} intra-shot corrections. From left to right, results without deliberate motion and reconstructions and errors in the presence of extreme motion for $Q=\{1,2,3\}$ repeats of a $R=2\times 2$ accelerated baseline scan.}
\label{fig:READ}

%% file: caption8.tex
\caption{Reconstruction results for pediatric cases with largest intra-scan degradations. \textbf{a)} Uncorrected; \textbf{b)} motion-corrected; and \textbf{c)} motion-corrected and regularized outlier segment rejection reconstructions. From left to right, results for the main families of sequences for structural brain imaging.}
\label{fig:REIN}

%% file: 5discussion.tex
\ifdefined\figures
\section{Discussion}
\else
\section*{Discussion}
\fi

\label{sec:DISC}

We have presented DISORDER, a retrospective framework for motion tolerant structural 3D k-space encoded brain imaging that combines optimized view orders with an improved aligned reconstruction. The proposed distributed and incoherent orders increase the motion sensitivity of the information sampled within a given time window which, provided a certain degree of redundancy, enables the resolvability of motion in the reconstruction. Conducted simulations have shown that reordering the k-space traversals introduces a significant boost in the ability to estimate the head pose and suppress motion artifacts. Tolerance to motion has been demonstrated in-vivo on a controlled experiment involving extreme and continuous motion throughout the examination as well as for the main families of sequences used for structural brain imaging by presenting the reconstruction results on the most challenging datasets from a pediatric cohort of $26$ subjects.

Although DISORDER is robust enough in its current form so as to be of practical interest for reliable structural brain MR examinations in non-compliant cohorts, with plans in our center to use it to progressively replace unnecessary sedation or anesthesia in pediatric and neonatal populations~\citep{Barton18}, it is obviously not free from limitations. First, data consistency may be affected by additional degrading factors. These include inaccuracies in sensitivities but also water-fat shifts, eddy currents or flow artifacts. In practice, applying fat suppression when possible, designing the tiles for adequate trade-offs between eddy currents and motion resolvability in bSSFP sequences, and adequate planning and scanning procedures are usually sufficient to address these issues. Differently, correction of non-rigid motion components would require an extension of the formulation. Although analogous methodologies~\citep{Loktyushin15} have shown potential in this context, a robust and efficient extension to non-rigid motion models will probably require a careful computational design. This may be particularly the case for high resolution applications, where both rigid and non-rigid motion become more important and challenging to correct~\citep{Budde14}. Moreover, coarse scale motion at ultra high field may require additional corrections of high order effects. Finally, in this manuscript we have restricted ourselves to uniform sampling, with further work required to generalize the incoherent and distributed orders and characterize motion correction and resolution retrieval when using variable densities.

In the in-vivo experiments of Fig.~\ref{fig:REIN} we have shown that inter-shot corrections can be sufficient in practical brain imaging scenarios requiring motion tolerance. Our underlying assumption is that the subject remains approximately still for a significant portion of the acquisition. In this case, inter-shot corrections are enough to reconcile the brain pose amongst the stable periods and data rejection can be applied to the transitions, again, provided that sampling is redundant enough. However, intra-shot corrections may become more important in challenging situations, as illustrated in Fig.~\ref{fig:READ}. Despite its computational limitations, our method is able to provide stable intra-shot estimates in the absence of motion while offering certain motion correction potential. Although a prior model for the temporal evolution of motion may aid in certain applications, in general, limitations arising from available computational resources and SNR per motion state are likely to complicate intra-shot tractability.

The situation may perhaps be different if using supervised learning strategies to inform the exploration of the motion parameter space. These may help to improve the spatio-temporal resolvability of motion by aiding the intra-shot corrections to find better motion solutions. Training may also help to enlarge the motion capture range at a given level of redundancy or decrease the required level of redundancy for a given motion capture range. Although direct learning of motion-corrected reconstructions could also be attempted, it is likely that, in many circumstances, better results will be obtained when concatenating learned reconstructions with model-based strategies, as recently suggested in~\cite{Haskell19}. Further integration of both approaches could be tackled, for instance, by incorporating the motion operator into the model-based learning framework in~\cite{Aggarwal19}, which may be effective in dealing with the residual penalties from g-factor amplification due to motion (see Fig.~\ref{fig:VARE}b). Thereby, future work will explore the opportunities for extending the ranges of motion resilience by supervised learning.

%% file: 6conclusion.tex
\ifdefined\figures
\section{Conclusion}
\else
\section*{Conclusion}
\fi

\label{sec:CONC}

We have proposed a simple modification of standard 3D Cartesian sequences for structural brain imaging, involving only a distributed and incoherent reordering of the sampled profiles, for high quality imaging in the presence of motion. Improved convergence has been demonstrated when using a separable nonlinear least squares formulation for joint motion estimation and reconstruction. Feasibility and conditions for inter- and intra-shot corrections have been characterized by simulations and in-vivo reconstructions under extreme motion. The DISORDER method has been successfully applied to replace sedation in a pediatric population scanned using common clinical examination protocols by combining inter-shot corrections with regularized outlier segment rejection reconstructions. Future work will focus on applying DISORDER to other cohorts and on strengthening its performance by integrating motion learning strategies.

%% file: acknowledgments.tex
\section*{Acknowledgments}

\ifdefined\figures This work received funding from the European Research Council under the European Union's Seventh Framework Programme (FP7/20072013/ERC, grant agreement no. [319456], dHCP project). The research was supported by the Wellcome/EPSRC Centre for Medical Engineering at King's College London [WT 203148/Z/16/Z]; the Medical Research Council [MR/K006355/1]; and the National Institute for Health Research (NIHR) Biomedical Research Centre based at Guy's and St Thomas' NHS Foundation Trust and King's College London. Jonathan O'Muircheartaigh is supported by a Sir Henry Dale Fellowship jointly funded by the Wellcome Trust and the Royal Society [206675/Z/17/Z]. The views expressed are those of the authors and not necessarily those of the NHS, the NIHR or the Department of Health. \fi The authors acknowledge the Department of Perinatal Imaging \& Health at King's College London.

%% file: 7support.tex
\ifdefined\figures
\section*{Supporting information}
\else
\section*{Supporting information}
\fi

\label{sec:SUMA}
Fig.~S1a collects the estimated motion traces for the cases in 
\ifdefined\external
 Fig.~8.
\else
 Fig.~\ref{fig:REIN}.
\fi
 Although no temporal regularization is used, all traces show periods of stability, which suggests accurate estimates at least in these periods. In Fig.~S1b the opacity of the traces is driven by the segment weights from the proposed outlier detection method. We observe that outliers generally correspond to main motion transients, in agreement to higher chances for intra-sweep degradation in these periods. 
\ifdefined\figures
\begin{figure*}[!htb]
\begin{center}
\hspace{8mm}\begin{overpic}[width=0.41\textwidth,draft=false]{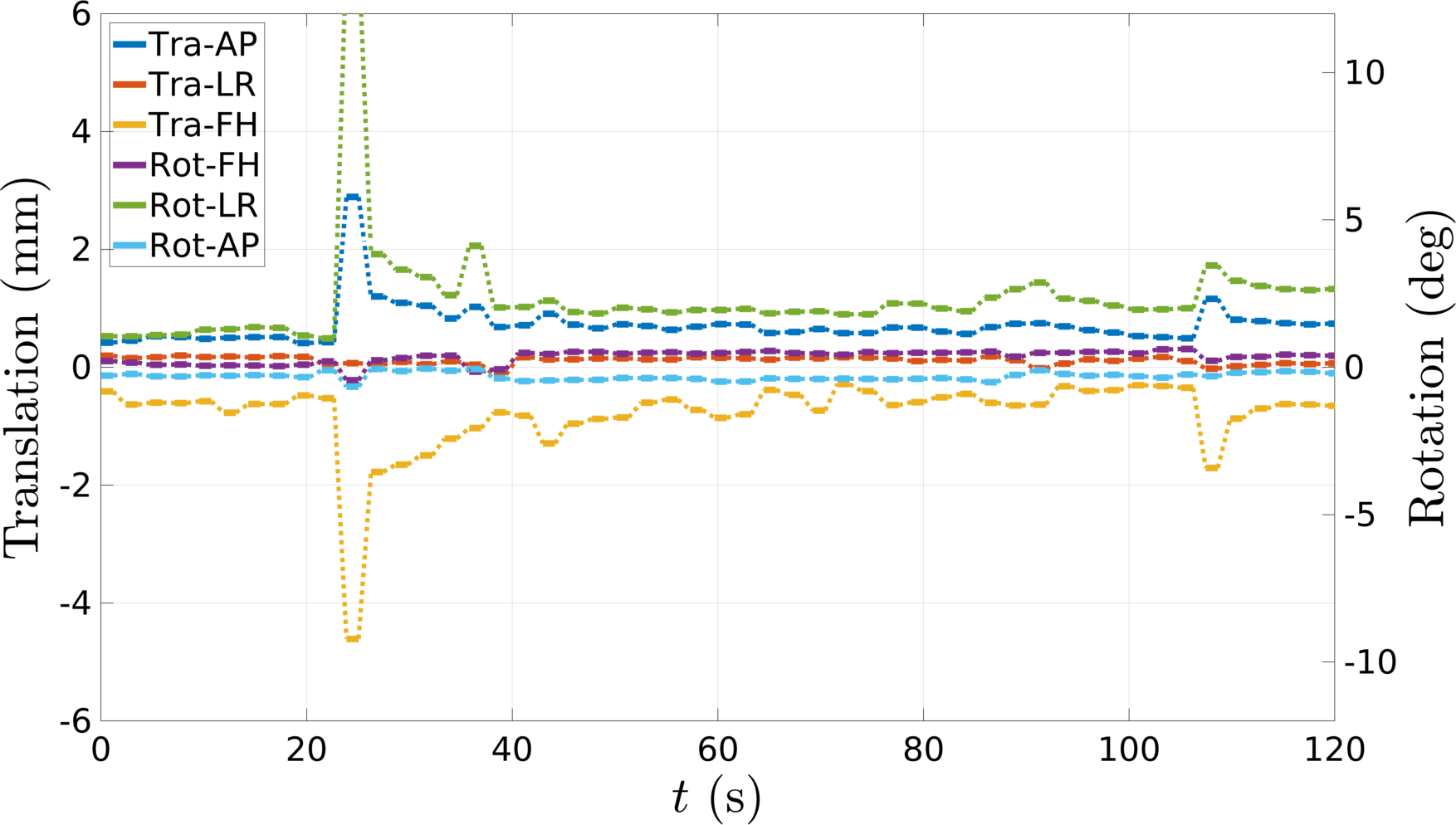}\put(-16,28){\makebox[0pt]{\Centerstack{MP-RAGE}}}\end{overpic}
\hspace{2mm}\begin{overpic}[width=0.41\textwidth,draft=false]{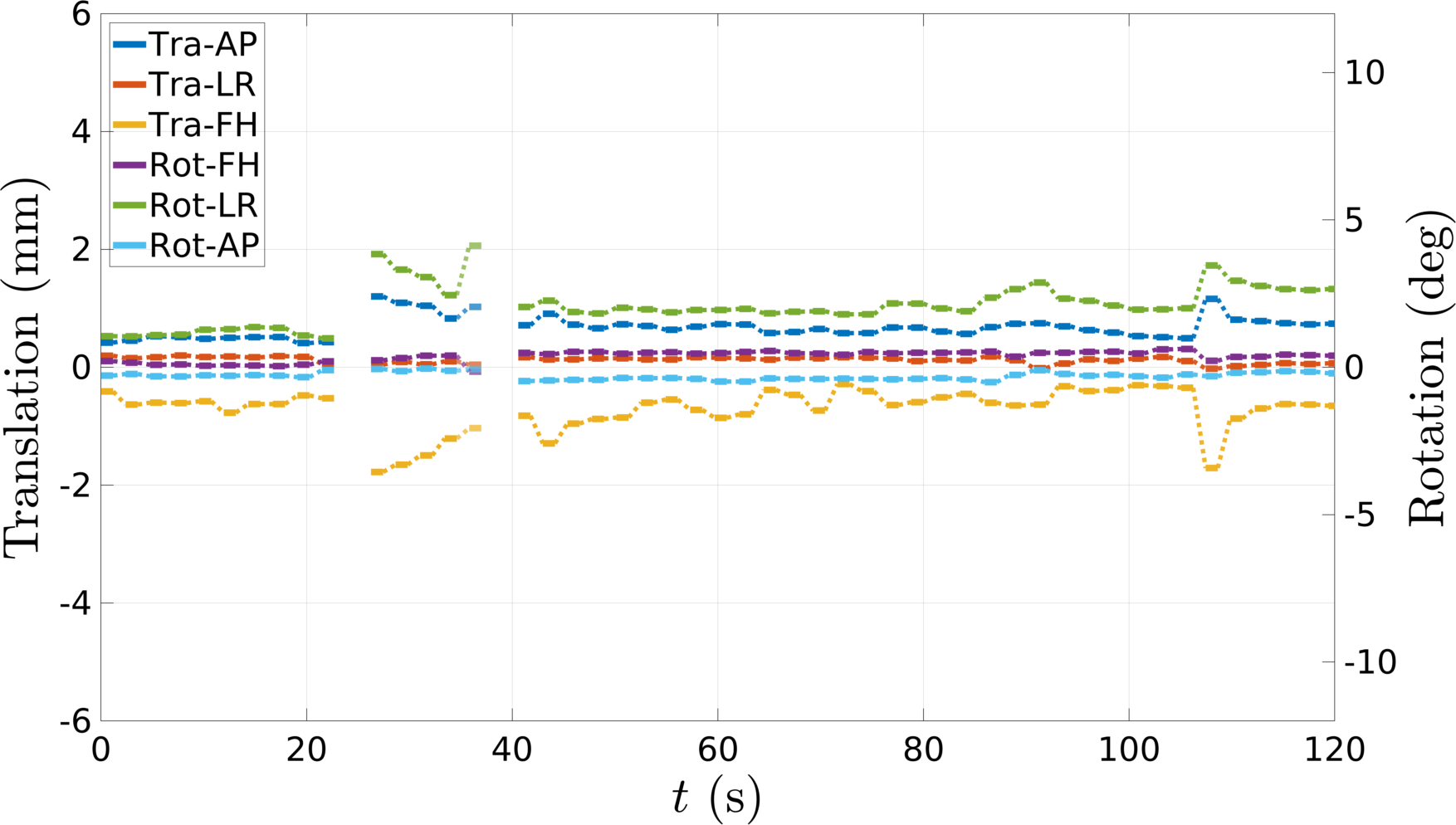}\end{overpic}\\
\hspace{8mm}\begin{overpic}[width=0.41\textwidth,draft=false]{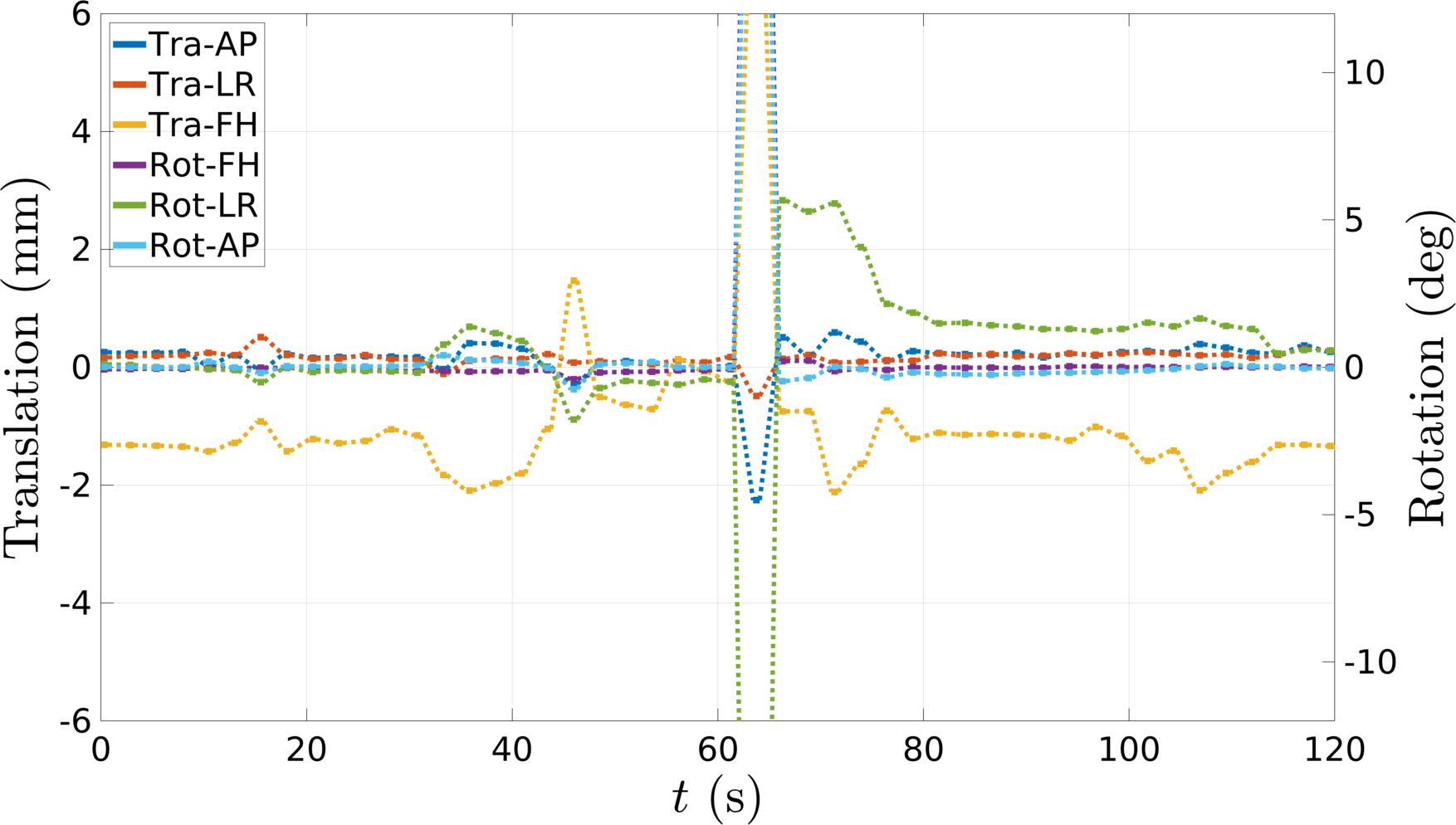}\put(-16,28){\makebox[0pt]{\Centerstack{TSE}}}\end{overpic}
\hspace{2mm}\begin{overpic}[width=0.41\textwidth,draft=false]{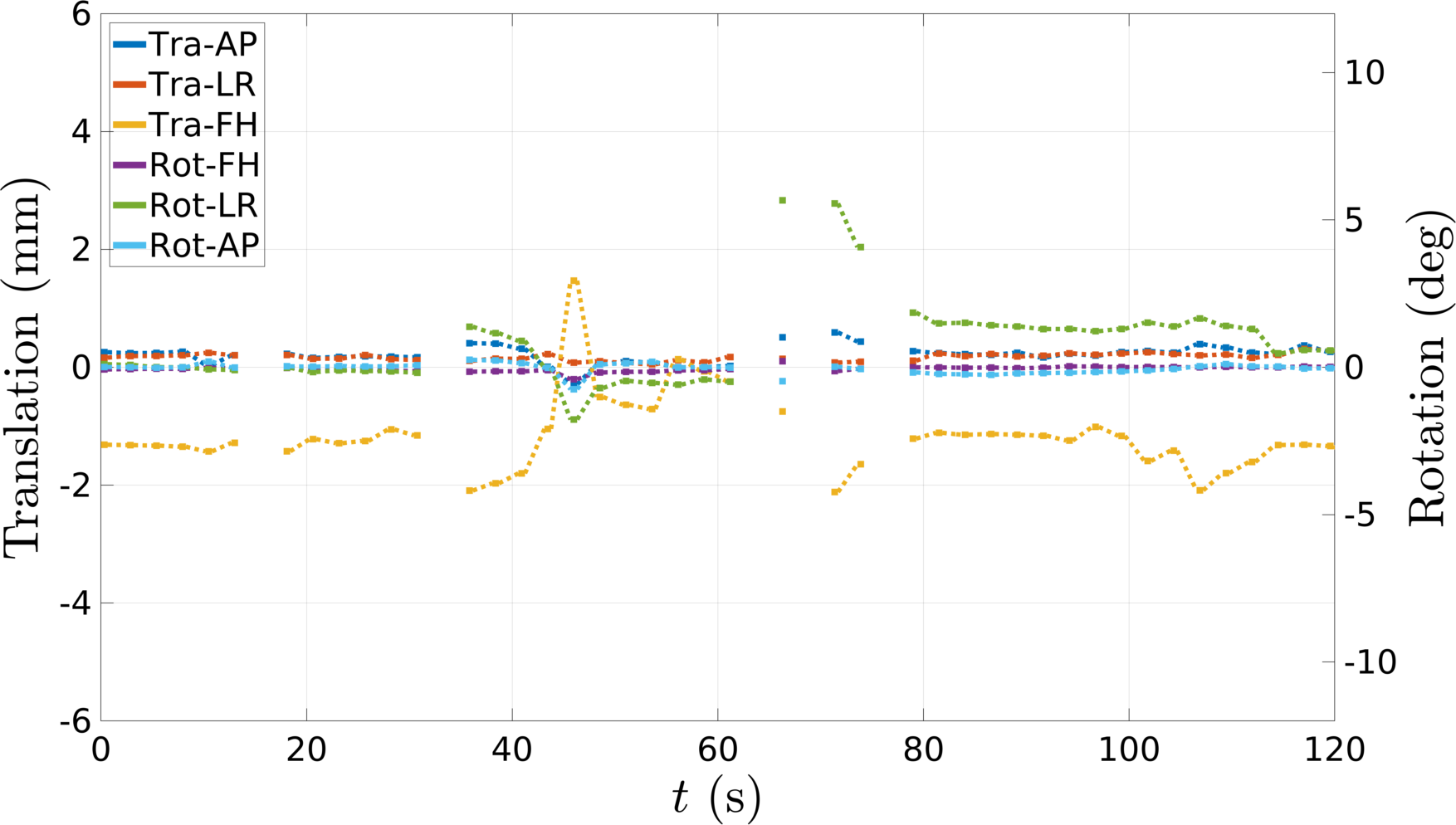}\end{overpic}\\
\hspace{8mm}\begin{overpic}[width=0.41\textwidth,draft=false]{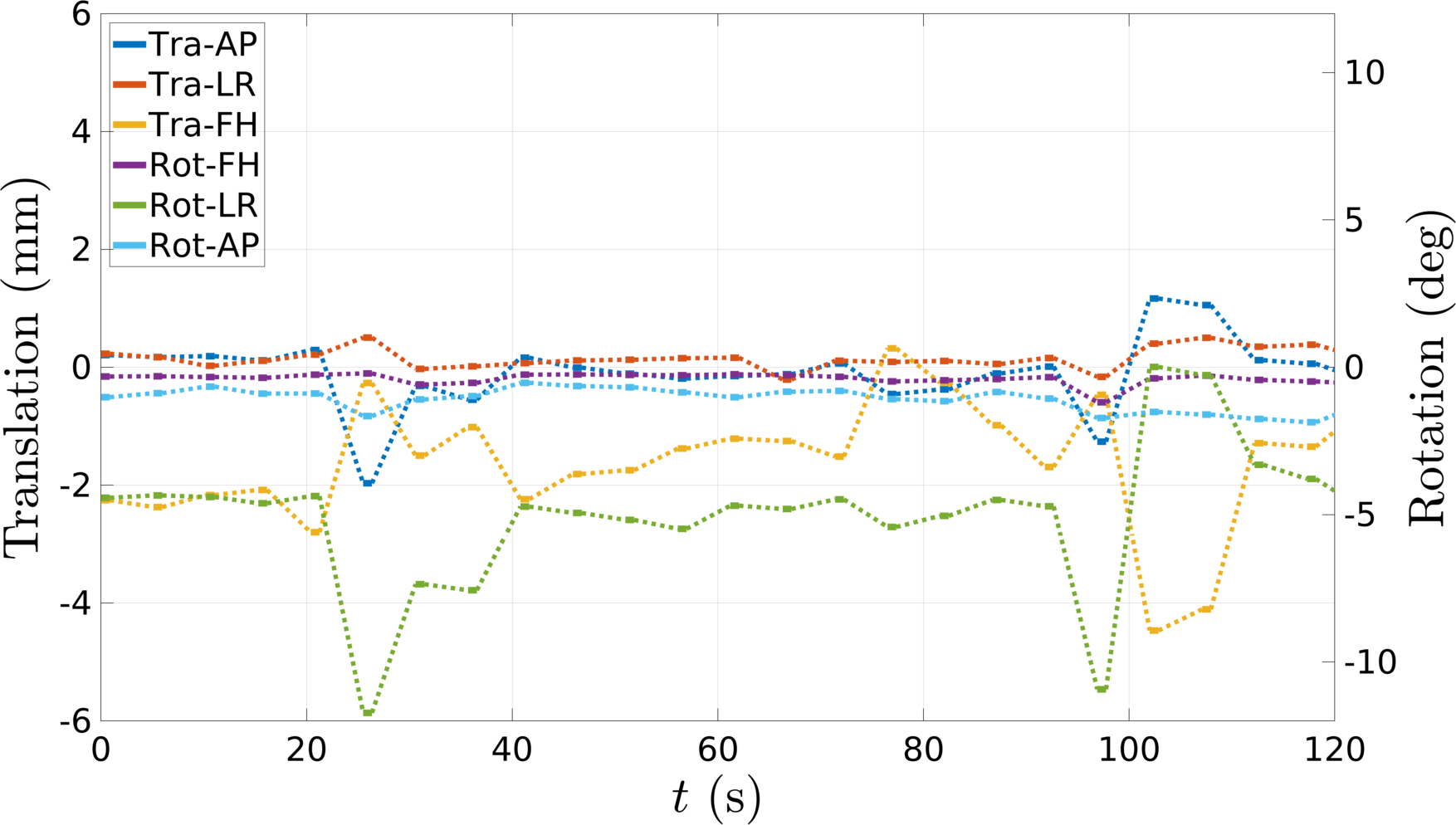}\put(-16,28){\makebox[0pt]{\Centerstack{FLAIR}}}\end{overpic}
\hspace{2mm}\begin{overpic}[width=0.41\textwidth,draft=false]{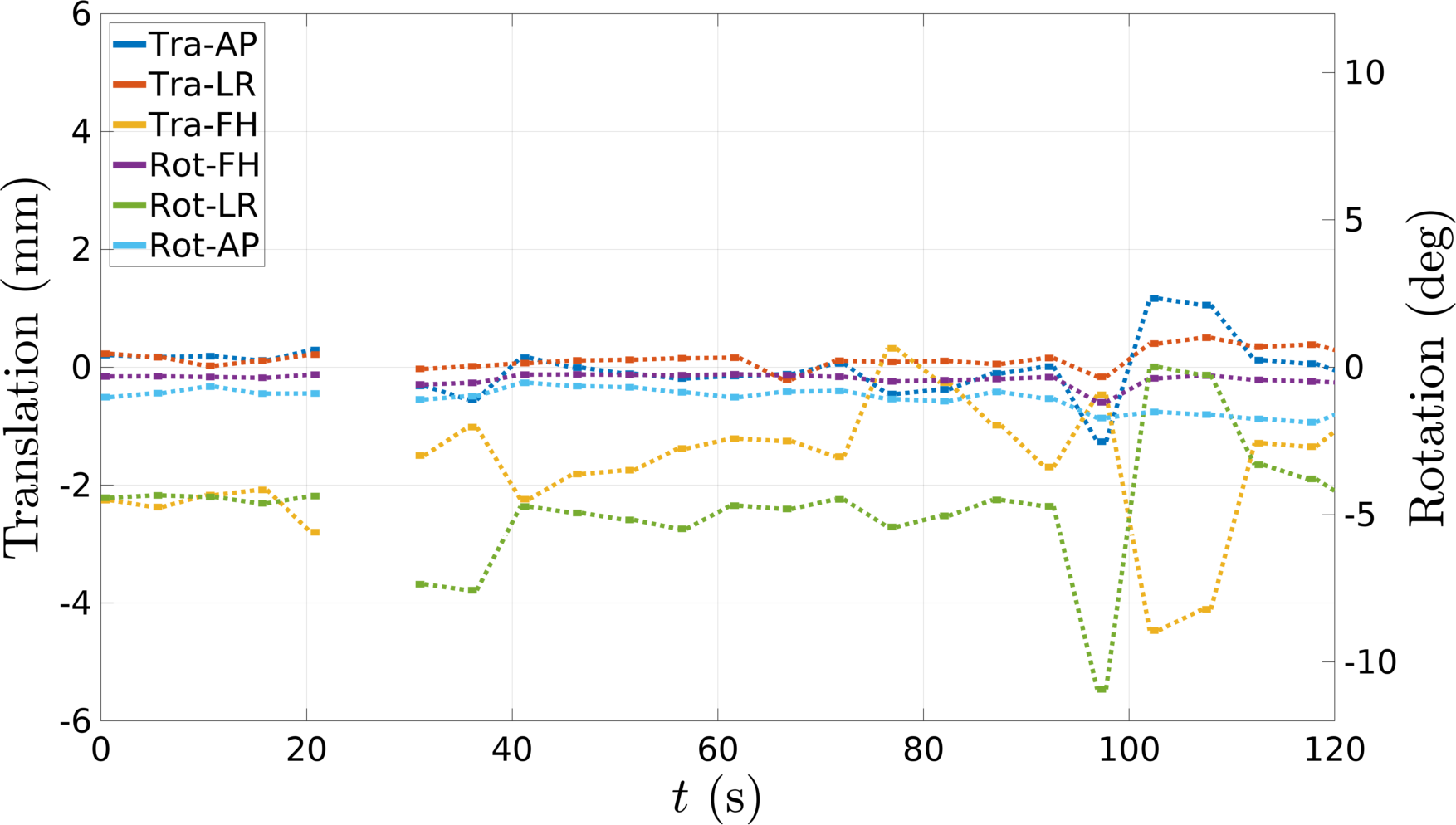}\end{overpic}\\
\hspace{8mm}\begin{overpic}[width=0.41\textwidth,draft=false]{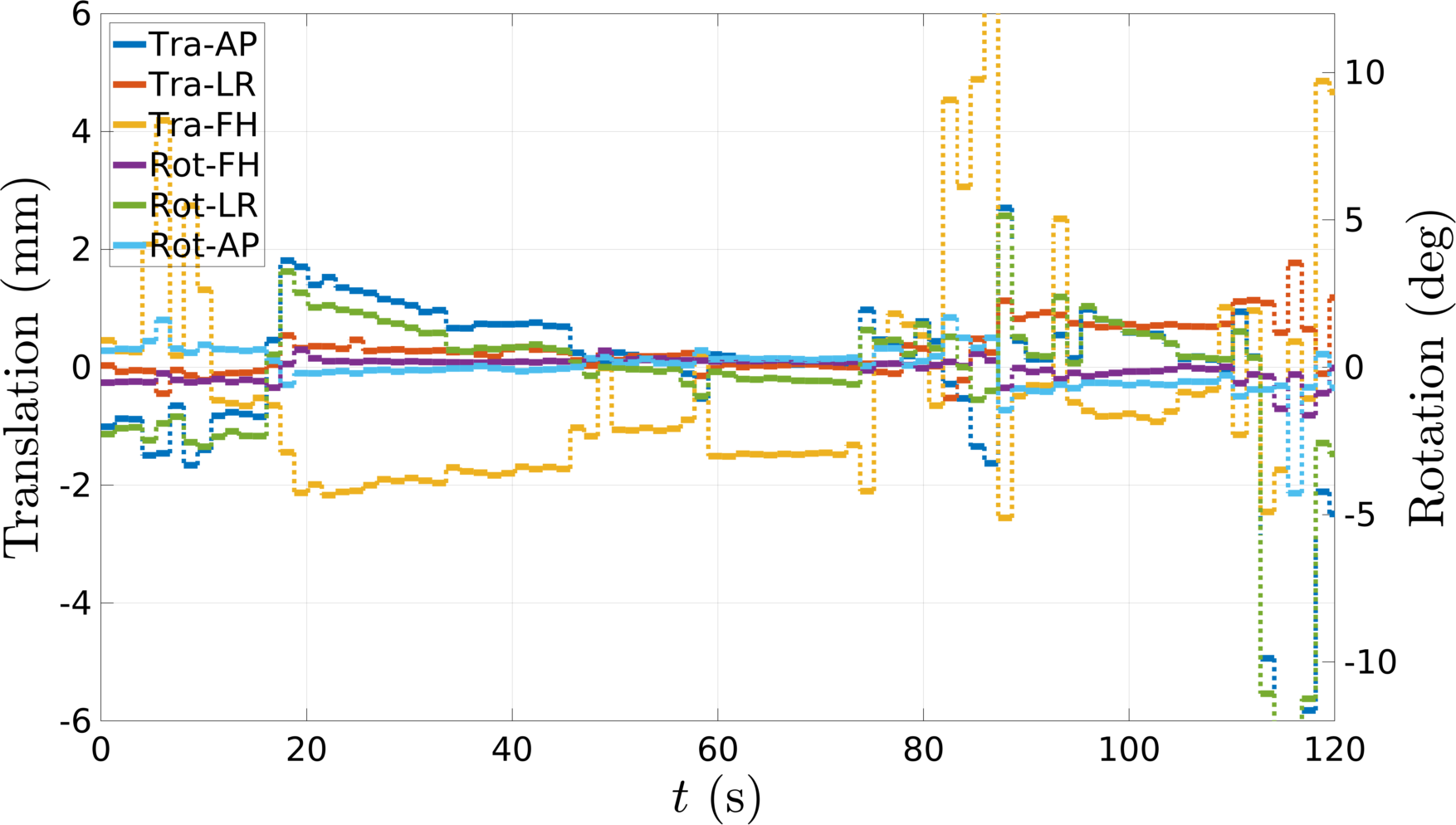}\put(-16,28){\makebox[0pt]{\Centerstack{SPGR}}}\end{overpic}
\hspace{2mm}\begin{overpic}[width=0.41\textwidth,draft=false]{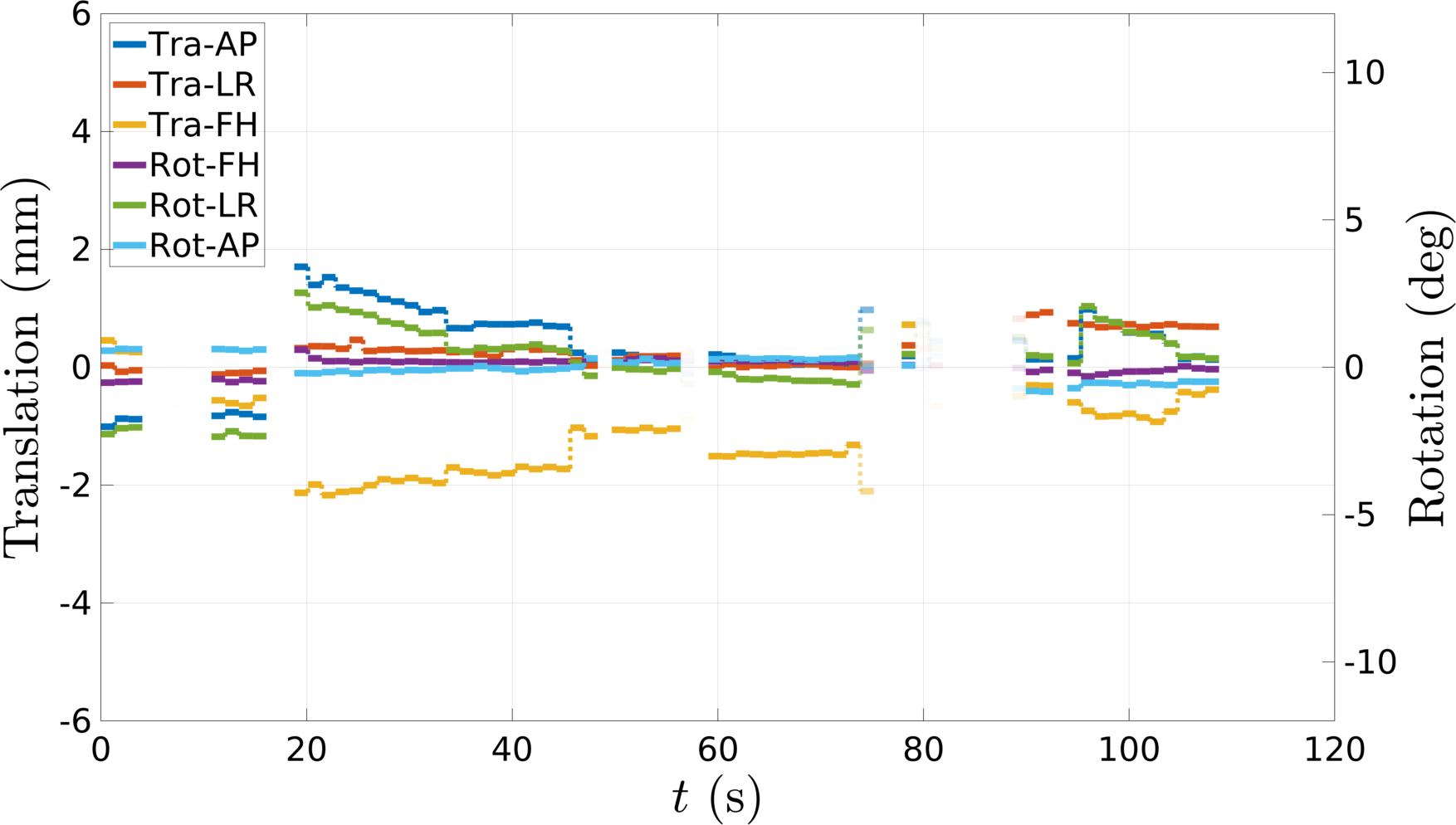}\end{overpic}\\
\hspace{8mm}\begin{overpic}[width=0.41\textwidth,draft=false]{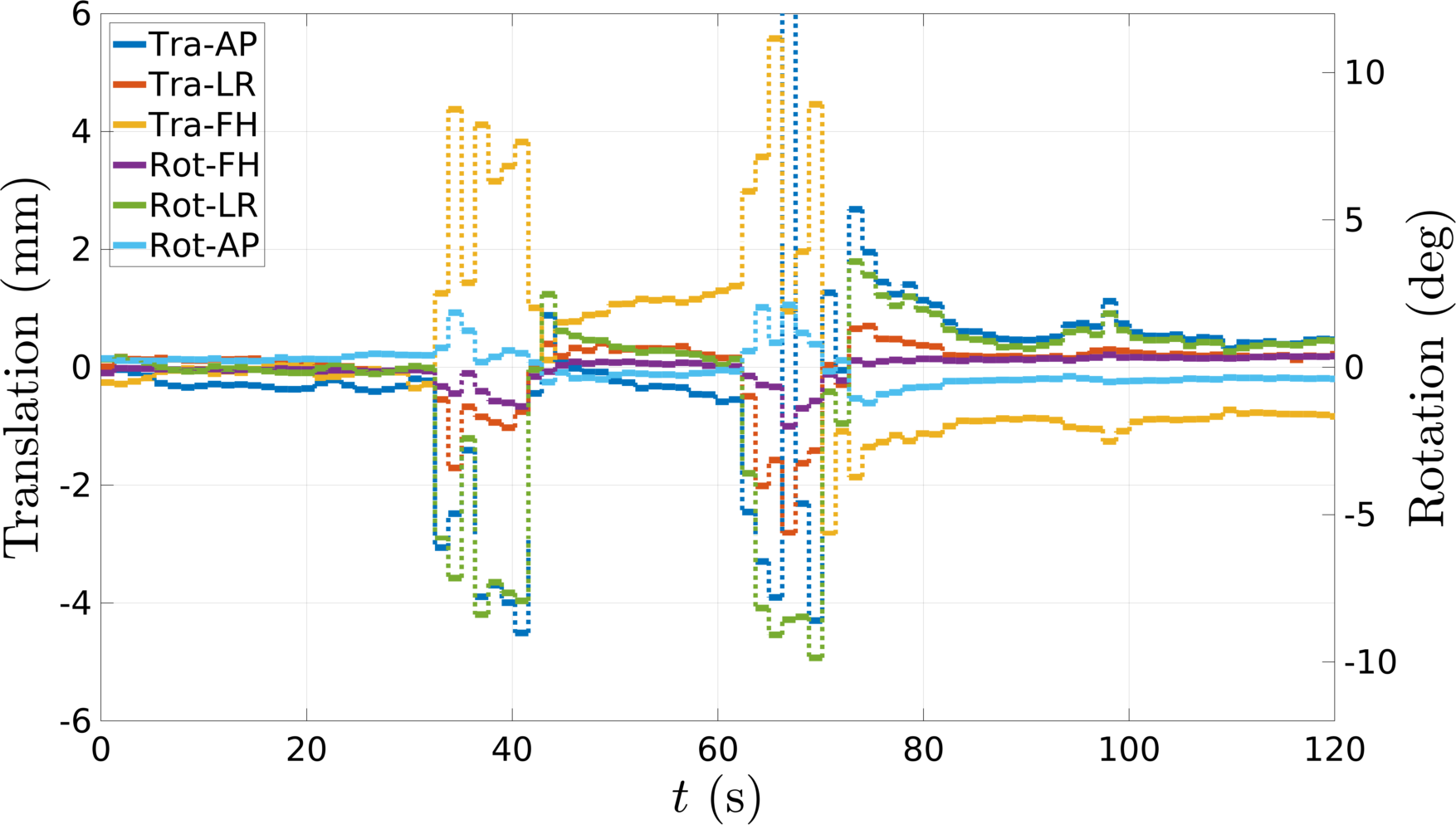}\put(-16,28){\makebox[0pt]{\Centerstack{BSSFP}}}\put(1,-6){\textbf{a)}}\end{overpic}
\hspace{2mm}\begin{overpic}[width=0.41\textwidth,draft=false]{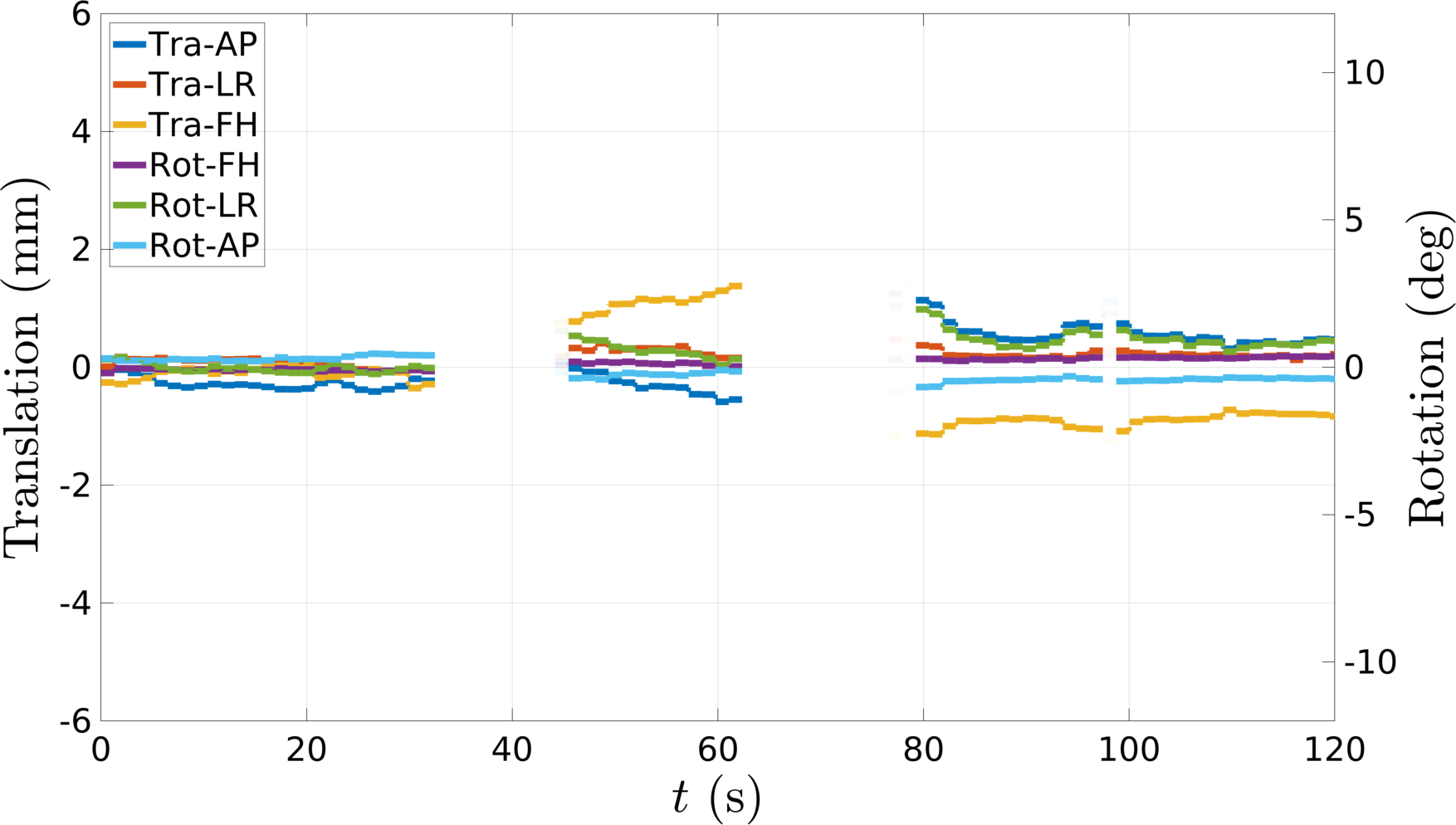}\put(1,-6){\textbf{b)}}\end{overpic}
\end{center}
\vspace{1mm}
\input{caption9}
\end{figure*}

%% file: caption9.tex
\textbf{Figure S1.} First $120\,\mbox{s}$ of estimated motion traces for pediatric cases with largest intra-scan degradations for each sequence. Left: original motion traces. Right: motion traces with segment opacity given by corresponding reliability $w$. Solid lines indicate data collection periods for each segment with dotted lines used to connect these. Tra and Rot in the legends refer respectively to Translation and Rotation.
\label{fig:MOTR}

%% file: DISORDER.bbl
\begin{thebibliography}{}

\bibitem[Adcock et~al., 2014]{Adcock14}
Adcock, B., Hansen, A., Roman, B., and Teschke, G. (2014).
\newblock Generalized sampling: Stable reconstructions, inverse problems and
  compressed sensing over the continuum.
\newblock {\em Adv. Imag. Electron Phys.}, 182(1):187--279.

\bibitem[Aggarwal et~al., 2019]{Aggarwal19}
Aggarwal, H.~K., Mani, M.~P., and Jacob, M. (2019).
\newblock {MoDL}: Model based deep learning architecture for inverse problems.
\newblock {\em IEEE Trans. Med. Imaging}, 38(2):394--405.

\bibitem[Aksoy et~al., 2012]{Aksoy12}
Aksoy, M., Forman, C., Straka, M., \c{C}ukur, T., Hornegger, J., and Bammer, R.
  (2012).
\newblock Hybrid prospective and retrospective head motion correction to
  mitigate cross-calibration errors.
\newblock {\em Magn. Reson. Med.}, 67(5):1237--1251.

\bibitem[Allison et~al., 2013]{Allison13}
Allison, M.~J., Ramani, S., and Fessler, J.~A. (2013).
\newblock Accelerated regularized estimation of {MR} coil sensitivities using
  augmented {L}agrangian methods.
\newblock {\em IEEE Trans. Med. Imaging}, 32(3):556--564.

\bibitem[{Anderson III} et~al., 2013]{Anderson13}
{Anderson III}, A.~G., Velikina, J., Block, W., Wieben, O., and Samsonov, A.
  (2013).
\newblock Adaptive retrospective correction of motion artifacts in cranial
  {MRI} with multicoil three-dimensional radial acquisitions.
\newblock {\em Magn. Reson. Med.}, 69(4):1094--1103.

\bibitem[Andre et~al., 2015]{Andre15}
Andre, J.~B., Bresnahan, B.~W., Mossa-Basha, M., Hoff, M.~N., Smith, C.~P.,
  Anzai, Y., and Cohen, W.~A. (2015).
\newblock Towards quantifying the prevalence, severity, and cost associated
  with patient motion during clinical {MR} examinations.
\newblock {\em J. Am. Coll. Radiol.}, 12(7):689--695.

\bibitem[Avron and Toledo, 2011]{Avron11}
Avron, H. and Toledo, S. (2011).
\newblock Randomized algorithms for estimating the trace of an implicit
  symmetric positive semi-definite matrix.
\newblock {\em J. Assoc. Comput. Mach.}, 58(2):8:1--8:17.

\bibitem[Bammer et~al., 2007]{Bammer07}
Bammer, R., Aksoy, M., and Liu, C. (2007).
\newblock Augmented generalized {SENSE} reconstruction to correct for rigid
  body motion.
\newblock {\em Magn. Reson. Med.}, 57(1):90--102.

\bibitem[Barton et~al., 2011]{Barton18}
Barton, K., Nickerson, J.~P., Higgins, T., and Williams, R.~K. (2011).
\newblock Pediatric anesthesia and neurotoxicity: what the radiologist needs to
  know.
\newblock {\em Pediatr. Radiol.}, 48(1):31--36.

\bibitem[Batchelor et~al., 2005]{Batchelor05}
Batchelor, P.~G., Atkinson, D., Irarrazaval, P., Hill, D. L.~G., Hajnal, J.,
  and Larkman, D. (2005).
\newblock Matrix description of general motion correction applied to multishot
  images.
\newblock {\em Magn. Reson. Med.}, 54(5):1273--1280.

\bibitem[Budde et~al., 2014]{Budde14}
Budde, J., Shajan, G., Scheffler, K., and Pohmann, R. (2014).
\newblock Ultra-high resolution imaging of the human brain using
  acquisition-weighted imaging at {9.4T}.
\newblock {\em NeuroImage}, 86(1):592--598.

\bibitem[Buehrer et~al., 2007]{Buehrer07}
Buehrer, M., Pruessmann, K.~P., Boesiger, P., and Kozerke, S. (2007).
\newblock Array compression for {MRI} with large coil arrays.
\newblock {\em Magn. Reson. Med.}, 57(6):1131--1139.

\bibitem[Busse et~al., 2008]{Busse08}
Busse, R.~F., Brau, A.~C., Vu, A., Michelich, C.~R., Bayram, E., Kijowski, R.,
  Reeder, S.~B., and Rowley, H.~A. (2008).
\newblock Effects of refocusing flip angle modulation and view ordering in {3D}
  fast spin echo.
\newblock {\em Magn. Reson. Med.}, 60(3):640--649.

\bibitem[Bydder et~al., 2002]{Bydder02}
Bydder, M., Larkman, D.~J., and Hajnal, J.~V. (2002).
\newblock Detection and elimination of motion artifacts by regeneration of
  $k$-space.
\newblock {\em Magn. Reson. Med.}, 47(4):677--686.

\bibitem[Cordero-Grande et~al., 2018]{Cordero-Grande18}
Cordero-Grande, L., Hughes, E.~J., Hutter, J., Price, A.~N., and Hajnal, J.~V.
  (2018).
\newblock Three-dimensional motion corrected sensitivity encoding
  reconstruction for multi-shot multi-slice {MRI}: Application to neonatal
  brain imaging.
\newblock {\em Magn. Reson. Med.}, 79(3):1365--1376.

\bibitem[Cordero-Grande et~al., 2016]{Cordero-Grande16}
Cordero-Grande, L., Teixeira, R. P. A.~G., Hughes, E.~J., Hutter, J., Price,
  A.~N., and Hajnal, J.~V. (2016).
\newblock Sensitivity encoding for aligned multishot magnetic resonance
  reconstruction.
\newblock {\em IEEE Trans. Comput. Imaging}, 2(3):266--280.

\bibitem[Fuderer et~al., 2004]{Fuderer04}
Fuderer, M., van~den Brink, J., and Jurrissen, M. (2004).
\newblock {SENSE} reconstruction using feed forward regularization.
\newblock In {\em 12th Proc. Intl. Soc. Mag. Reson. Med.}, page 2130, Kyoto,
  Japan.

\bibitem[Gallichan and Marques, 2017]{Gallichan17}
Gallichan, D. and Marques, J.~P. (2017).
\newblock Optimizing the acceleration and resolution of three-dimensional fat
  image navigators for high-resolution motion correction at {7T}.
\newblock {\em Magn. Reson. Imaging}, 77(2):547--558.

\bibitem[Gan et~al., 2018]{Gan18}
Gan, M., Chen, C. L.~P., Chen, G.-Y., and Chen, L. (2018).
\newblock On some separated algorithms for separable nonlinear least squares
  problems.
\newblock {\em IEEE Trans. Cybern.}, 48(10):2866--2874.

\bibitem[Godenschweger et~al., 2016]{Godenschweger16}
Godenschweger, F., K{\"a}gebein, U., Stucht, D., Yarach, U., Sciarra, A.,
  Yakupov, R., L{\"u}sebrink, F., Schulze, P., and Speck, O. (2016).
\newblock Motion correction in {MRI} of the brain.
\newblock {\em Phys. Med. Biol.}, 61(5):R32--R56.

\bibitem[Haskell et~al., 2019]{Haskell19}
Haskell, M.~W., Cauley, S.~F., Bilgic, B., Hossbach, J., Splitthoff, D.~N.,
  Pfeuffer, J., Setsompop, K., and Wald, L.~L. (2019).
\newblock Network accelerated motion estimation and reduction ({NAMER}):
  Convolutional neural network guided retrospective motion correction using a
  separable motion model.
\newblock {\em Magn. Reson. Med.}, 82(4):1452--1461.

\bibitem[Herring et~al., 2018]{Herring18}
Herring, J.~L., Nagy, J.~G., and Ruthotto, L. (2018).
\newblock {LAP}: A linearize and project method for solving inverse problems
  with coupled variables.
\newblock {\em Sampling Theory Signal Image Process.}, 17(2):127--151.

\bibitem[Johnson et~al., 2016]{Johnson16}
Johnson, P.~M., Liu, J., Wade, T., Tavallaei, M.~A., and Drangova, M. (2016).
\newblock Retrospective {3D} motion correction using spherical navigator
  echoes.
\newblock {\em Magn. Reson. Imaging}, 34(9):1274--1282.

\bibitem[Kutyniok et~al., 2012]{Kutyniok12}
Kutyniok, G., Lemvig, J., and Lim, W.-Q. (2012).
\newblock Optimally sparse approximations of {3D} functions by compactly
  supported shearlet frames.
\newblock {\em SIAM J. Math. Anal.}, 44(4):2962--3017.

\bibitem[Kutyniok et~al., 2016]{Kutyniok16}
Kutyniok, G., Lim, W.-Q., and Reisenhofer, R. (2016).
\newblock {ShearLab 3D}: Faithful digital shearlet transforms based on
  compactly supported shearlets.
\newblock {\em ACM Trans. Math. Softw.}, 42(1):5:1--5:42.

\bibitem[Li and Zhong, 2018]{Li18}
Li, C.-J. and Zhong, Y.-J. (2018).
\newblock A pseudo-heuristic parameter selection rule for $l^1$-regularized
  minimization problems.
\newblock {\em J. Comput. Appl. Math.}, 333:1--19.

\bibitem[Loktyushin et~al., 2013]{Loktyushin13}
Loktyushin, A., Nickisch, H., Pohmann, R., and Sch{\"o}lkopf, B. (2013).
\newblock Blind retrospective motion correction of {MR} images.
\newblock {\em Magn. Reson. Med.}, 70(6):1608--1618.

\bibitem[Loktyushin et~al., 2015]{Loktyushin15}
Loktyushin, A., Nickisch, H., Pohmann, R., and Sch{\"o}lkopf, B. (2015).
\newblock Blind multirigid retrospective motion correction of {MR} images.
\newblock {\em Magn. Reson. Med.}, 73(4):1457--1468.

\bibitem[Maclaren et~al., 2013]{Maclaren13}
Maclaren, J., Herbst, M., Speck, O., and Zaitsev, M. (2013).
\newblock Prospective motion correction in brain imaging: A review.
\newblock {\em Magn. Reson. Med.}, 69(3):621--636.

\bibitem[Odille et~al., 2008]{Odille08}
Odille, F., Vuissoz, P.-A., Marie, P.-Y., and Felblinger, J. (2008).
\newblock Generalized reconstruction by inversion of coupled systems ({GRICS})
  applied to free-breathing {MRI}.
\newblock {\em Magn. Reson. Med.}, 60(1):146--157.

\bibitem[Pipe et~al., 2014]{Pipe14}
Pipe, J.~G., Gibbs, W.~N., Li, Z., Karis, J.~P., Schar, M., and Zwart, N.~R.
  (2014).
\newblock Revised motion estimation algorithm for {PROPELLER MRI}.
\newblock {\em Magn. Reson. Med.}, 72(2):430--437.

\bibitem[Pruessmann et~al., 2001]{Pruessmann01}
Pruessmann, K.~P., Weiger, M., B{\"o}rnert, P., and Boesiger, P. (2001).
\newblock Advances in sensitivity encoding with arbitrary k-space trajectories.
\newblock {\em Magn. Reson. Med.}, 46(4):638--651.

\bibitem[Pruessmann et~al., 1999]{Pruessmann99}
Pruessmann, K.~P., Weiger, M., Scheidegger, M.~B., and Boesiger, P. (1999).
\newblock {SENSE}: sensitivity encoding for fast {MRI}.
\newblock {\em Magn. Reson. Med.}, 42(5):952--962.

\bibitem[Ruano et~al., 1991]{Ruano91}
Ruano, A. E.~B., Jones, D.~I., and Fleming, P.~J. (1991).
\newblock A new formulation of the learning problem of a neural network
  controller.
\newblock In {\em Proc. 30th {IEEE} Conf. Decision Control}, pages 865--866,
  Brighton, England.

\bibitem[Samsonov et~al., 2010]{Samsonov10}
Samsonov, A.~A., Velikina, J., Jung, Y., Kholmovski, E.~G., Johnson, C.~R., and
  Block, W.~F. (2010).
\newblock {POCS}-enhanced correction of motion artifacts in parallel {MRI}.
\newblock {\em Magn. Reson. Med.}, 63(4):1104--1110.

\bibitem[Schulz et~al., 2012]{Schulz12}
Schulz, J., Siegert, T., Reimer, E., Labadie, C., Maclaren, J., Herbst, M.,
  Zaitsev, M., and Turner, R. (2012).
\newblock An embedded optical tracking system for motion-corrected magnetic
  resonance imaging at {7T}.
\newblock {\em Magn. Reson. Mater. Phy.}, 25(6):443--453.

\bibitem[Tisdall et~al., 2016]{Tisdall16}
Tisdall, M.~D., Reuter, M., Qureshi, A., Buckner, R.~L., Fischl, B., and
  van~der Kouwe, A.~J. (2016).
\newblock Prospective motion correction with volumetric navigators ({vNavs})
  reduces the bias and variance in brain morphometry induced by subject motion.
\newblock {\em NeuroImage}, 127:11--22.

\bibitem[Tsao et~al., 2005]{Tsao05}
Tsao, J., Kozerke, S., Boesiger, P., and Pruessmann, K.~P. (2005).
\newblock Optimizing spatiotemporal sampling for k-t {BLAST} and k-t {SENSE}:
  Application to high-resolution real-time cardiac steady-state free
  precession.
\newblock {\em Magn. Reson. Med.}, 53(6):1372--1382.

\bibitem[Unser et~al., 1993]{Unser93}
Unser, M., Aldroubi, A., and Gerfen, C.~R. (1993).
\newblock A multiresolution image registration procedure using spline pyramids.
\newblock In {\em Proc. SPIE, Wavelet Applications in Signal and Image
  Processing}, volume 2034, pages 160--170.

\bibitem[Unser et~al., 1995]{Unser95}
Unser, M., Th\'evenaz, P., and Yaroslavsky, L. (1995).
\newblock Convolution-based interpolation for fast, high-quality rotation of
  images.
\newblock {\em IEEE Trans. Image Process.}, 4(10):1375--1381.

\bibitem[Voronin and Daubechies, 2017]{Voronin17}
Voronin, S. and Daubechies, I. (2017).
\newblock {\em Functional Analysis, Harmonic Analysis, and Image Processing},
  volume 693 of {\em Contemp. Math.}, chapter An iteratively reweighted least
  squares algorithm for sparse regularization, pages 391--411.
\newblock Am. Math. Soc.

\bibitem[Zaitsev et~al., 2015]{Zaitsev15}
Zaitsev, M., Maclaren, J., and Herbst, M. (2015).
\newblock Motion artifacts in {MRI}: A complex problem with many partial
  solutions.
\newblock {\em J. Magn. Reson. Imaging}, 42(4):887--901.

\end{thebibliography}
